\documentclass[aps,prd,twocolumn,superscriptaddress,showpacs,floatfix]{revtex4}
\usepackage[dvips]{graphicx}

\newcommand{\lsim}{\lower0.6ex\vbox{\hbox{$ \buildrel{\textstyle <}\over{\sim}\ $}}}
\newcommand{\gsim}{\lower0.6ex\vbox{\hbox{$ \buildrel{\textstyle >}\over{\sim}\ $}}}
\newcommand{\beq}{\begin{equation}}
\newcommand{\eeq}{\end{equation}}

\newcommand{\hMsun}{\ h^{-1}\mathrm{M}_{\odot}}
\newcommand{\hMpc}{\ h^{-1}\mathrm{Mpc}}

\newcommand{\Omegam}{\Omega_{\mathrm{M}}}

\newcommand{\Omegade}{\Omega_{\mathrm{DE}}}

\newcommand{\omegam}{\omega_{\mathrm{M}}}
\newcommand{\omegab}{\omega_{\mathrm{B}}}
\newcommand{\ns}{n_{\mathrm{s}}}
\newcommand{\dr}{\Delta_{\mathcal{R}}^{2}}
\newcommand{\wzero}{w_{0}}
\newcommand{\wa}{w_{\mathrm{a}}}
\newcommand{\apiv}{a_{\mathrm{p}}}
\newcommand{\wpiv}{w_{\mathrm{p}}}
\newcommand{\fom}{\mathcal{F}~}

\newcommand{\rhomean}{\rho_{\mathrm{M}}}

\newcommand{\Poneh}{P_{1\mathrm{H}}}
\newcommand{\Ptwoh}{P_{2\mathrm{H}}}
\newcommand{\Rvir}{R_{200\mathrm{m}}}

\newcommand{\bh}{b_{\mathrm{h}}}

\newcommand{\zpiv}{z_{\mathrm{piv}}}
\newcommand{\mpiv}{m_{\mathrm{piv}}}
\newcommand{\cpiv}{c_{\mathrm{piv}}}

\newcommand{\Pktom}[2]{P_{\kappa}^{\mathrm{#1}\mathrm{#2}}}
\newcommand{\Pkobs}[2]{\bar{P}_{\kappa}^{\mathrm{#1}\mathrm{#2}}}

\newcommand{\ellmax}{\ell_{\mathrm{max}}}
\newcommand{\wlw}[1]{W_{\mathrm{#1}}}
\newcommand{\Da}{D_{\mathrm{A}}}

\newcommand{\ntomo}{N_{\mathrm{TOM}}}

\newcommand{\zp}{z_{\mathrm{p}}}

\newcommand{\fsky}{f_{\mathrm{sky}}}
\newcommand{\gi}{\langle \gamma^2 \rangle}

\newcommand{\dd}{\mathrm{d}}

\begin{document}


\title{
Self Calibration of Tomographic Weak Lensing for the Physics of Baryons to Constrain Dark Energy
}

\author{Andrew R. Zentner}
\affiliation{
Department of Physics and Astronomy, University of Pittsburgh, 
Pittsburgh, PA 15260
}
\email{
zentner@pitt.edu
}
\affiliation{
Kavli Institute for Cosmological Physics and Department of Astronomy and Astrophysics,
The University of Chicago, Chicago, IL 60637
}
\affiliation{
The Enrico Fermi Institute, The University of Chicago, Chicago, IL 60637
}
\author{Douglas H. Rudd}
\affiliation{
Kavli Institute for Cosmological Physics and Department of Astronomy and Astrophysics,
The University of Chicago, Chicago, IL 60637
}
\author{Wayne Hu}
\affiliation{
Kavli Institute for Cosmological Physics and Department of Astronomy and Astrophysics,
The University of Chicago, Chicago, IL 60637
}
\affiliation{
The Enrico Fermi Institute, The University of Chicago, Chicago, IL 60637
}

\date{\today}


\begin{abstract}

Recent numerical studies indicate that uncertainties in the treatment of baryonic physics
can affect predictions for weak lensing shear 
power spectra at a level that is significant for several forthcoming surveys 
such as the Dark Energy Survey (DES), the 
SuperNova/Acceleration Probe (SNAP), and the Large Synoptic Survey Telescope (LSST).
Correspondingly, we show that baryonic effects can significantly bias   
dark energy parameter measurements.  Elimination of 
such potential biases by neglecting information in multipoles
beyond several hundred leads to weaker parameter constraints by
a factor of $\sim 2-3$ compared with using information out 
to multipoles of several thousand.  
Fortunately, the same numerical studies that explore the influence 
of baryons indicate that they primarily affect power spectra by altering halo 
structure through the relation between halo mass and mean effective halo 
concentration.  We explore the ability of future weak lensing surveys to constrain 
both the internal structures of halos and the properties of the dark energy 
simultaneously as a first step toward self calibrating for the physics of 
baryons.  In this approach, parameter biases are 
greatly reduced and no parameter constraint is degraded by 
more than $\sim 40\%$ in the case of LSST or $30\%$ in the cases 
of SNAP or DES.  Modest prior knowledge of the halo concentration relation 
and its redshift evolution greatly improves even these forecasts.  
In addition, we find that these surveys can constrain effective halo 
concentrations themselves usefully with shear power spectra alone.  In the 
most restrictive case of a power-law relation for halo concentration 
as a function of mass and redshift, 
the concentrations of halos of mass $m \sim 10^{14} \hMsun$ 
at $z \sim 0.2$ can be constrained to better than $10\%$.  
Our results suggest that inferring dark energy parameters through shear spectra 
can be made robust to baryonic physics and that this procedure may even provide 
useful constraints on galaxy formation models.

\end{abstract}




\maketitle


\section{Introduction}
\label{section:introduction}

An ever-expanding set of observational data indicate that roughly $\sim 75\%$ of the 
energy in the universe is in the form of {\em dark energy} whose negative pressure 
drives an accelerated cosmological expansion 
(e.g.,~\cite{riess_etal98,perlmutter_etal99,tegmark_etal04,riess_etal04,
eisenstein_etal05,spergel_etal07,tegmark_etal06,astier_etal06,wood-vasey_etal07}). 
Determining the nature of the dark energy is one of the most profound problems facing 
physicists and astronomers.  Numerous contemporary and forthcoming experiments aim to shed light on 
the dark energy problem using a variety of techniques.   One of the most promising 
probes of dark energy is weak gravitational lensing.  Forthcoming weak lensing 
surveys expect to measure the statistics of the matter density fluctuation field 
with exquisite precision.  Dark energy affects both the rate of structure growth and 
the relative angular diameter distances between 
different redshifts and so these precise determinations 
of the matter fluctuation spectrum bring the promise of stringent constraints on 
dark energy parameters \cite{hu_tegmark99,hu99,huterer02,heavens03,refregier03,
refregier_etal04,song_knox04,takada_jain04,takada_white04,dodelson_zhang05,
albrecht_etal06,zhan06}.

Of course, exploiting precise measurements of fluctuations in 
the density field to constrain dark 
energy requires precise predictions for these inhomogeneities.  
Maximizing the constraining power of 
future surveys in this regard demands that predictions 
for the linear and nonlinear matter power spectra be 
accurate at the percent level \cite{huterer_takada05,huterer_etal06}.  
This remains a challenge for numerical simulations even in the absence of baryonic physics 
\cite{heitmann_etal05}, but this challenge should yield to unrelenting increases in computational 
power.  Perhaps more important is the treatment of the baryonic component of the universe.  
Early analytic studies indicated that baryonic processes may have a significant influence on 
shear power spectra \cite{white04,zhan_knox04}.  More recently, \citet{jing_etal06} and \citet{rudd_etal07}  
showed that numerical simulations that 
include baryonic processes predict matter power spectra that differ 
from those that result from dissipationless $N$-body calculations at a level that 
exceeds the precision of various forthcoming experiments.  
The evolution of the baryonic component of the universe cannot be modeled directly 
with current computational limitations and this circumstance is unlikely to 
yield to increases in computational power in the 
foreseeable future.  As a consequence, all calculations 
rely on effective models that attempt to approximate the net, large-scale influences 
of numerous processes that occur on scales far below 
the numerical resolution that may be attained with any simulation.  
The implication of Refs.~\cite{jing_etal06,rudd_etal07} is that the ability of weak lensing surveys 
to constrain dark energy parameters may be severely limited by this inability to account 
for baryonic processes reliably.

The goal of this study is to investigate this implication.  
In particular, we aim to determine if 
measurements of shear power spectra (including tomography with photometric redshifts) contain 
sufficient information to constrain both the influence of baryons and 
the properties of the dark energy.  
\citet{rudd_etal07} found that the influence of baryons 
is due almost entirely to modifications in the matter distribution within dark matter halos.  
We model the influence of baryons in this way, so our calculation is tantamount to 
determining the ability of future surveys to constrain a physically-motivated model 
for the density profiles of dark matter halos and dark energy parameters simultaneously.  
Naturally, gravitational lensing is sensitive to all matter along the lines of sight to 
sources.  In the following we refer to the distribution of mass within halos for simplicity, 
but we do not intend to restrict our consideration to the dark matter.  We use this 
terminology as a shorthand to refer to the total mass distribution of the 
composite dark matter and baryonic systems within halo virial radii.

Our results suggest that forthcoming surveys will have sufficient information to calibrate 
for the influence of baryons on the matter power spectrum using weak lensing shear power 
spectra alone, with minimal degradation in the constraints on dark energy parameters.  In 
fact, we study models that are quite general so our results suggest that such surveys 
can also calibrate additional systematics that may afflict a similar range of scales.  
Of course, one need not regard halo structure strictly as a nuisance.  
An interesting concomitant outcome of this calibration program is that 
weak lensing surveys will provide constraints on halo 
structure that complement constraints obtained via alternative means 
(e.g.,~Ref.~\cite{sheldon_etal04,lokas_etal06,mandelbaum_etal06,buote_etal06,
bolton_etal06,johnston_etal07,gavazzi_etal07,comerford_natarajan07}).  This byproduct of dark 
energy studies extends the scientific reach of future surveys as it may prove useful 
to inform models of the formation of galaxies and galaxy clusters.

Alongside this success, we emphasize that our results must be considered 
with some circumspection.  Though it appears to be the dominant effect, 
we do not expect that the influence of baryons is strictly limited to 
the internal structures of halos.  In our study 
we also neglect all systematics other than baryonic physics, such as intrinsic 
alignments \cite{hirata_seljak04,heymans_etal04,mandelbaum_etal06b,bridle_king07}.  
We consider this exercise to be a 
proof of concept indicating that self calibration is achievable 
at minimal cost and that weak lensing tomography may constrain halo 
profiles at interesting levels.

There are several papers in the literature that are closely related to our work.  We have 
already mentioned that both \citet{white04} and \citet{zhan_knox04} 
studied the effects of baryons on convergence spectra using analytic models and 
indicated that they will likely be non-negligible, at least over some range of scales.  
Though prescient, both of these studies 
are based on phenomenological models that do not treat all of the effects of baryons 
self consistently, while the study of \citet{zhan_knox04} 
focuses on the influence of hot baryons and neglect 
baryonic cooling.  The result of these studies is that baryonic processes influence lensing 
observables on scales near $\ell \sim 10^3$ at a level below that reported in 
Refs.~\cite{jing_etal06,rudd_etal07}.  Furthermore, it is not entirely surprising that 
the information within forthcoming lensing surveys suffices to calibrate additional physics, 
such as the influence of baryonic processes.  \citet{huterer_etal06} studied self calibration 
of multiplicative and additive observational systematics and found that these could be 
calibrated successfully at a relatively low cost.  Self calibration of systematics, including 
intrinsic alignment effects is also discussed briefly in the appendices of the report of 
the Dark Energy Task Force (DETF, Ref.~\cite{detf}) and such self calibration is 
included in their analyses.

We describe our methods in the following section.  
We begin with a review of weak lensing 
tomography.  Next, we summarize the salient features of the 
halo model for nonlinear clustering.  We then review the Fisher matrix formalism 
for assessing the constraining power of forthcoming experiments and mention some of the specifics 
of our implementation, including our fiducial model and priors.  
We close the second section with a variety of models that we use to incorporate parameterized 
baryonic physics into our calculations.

In Section~\ref{section:results}, we present the results of our study.  We open with a 
review of forecasts for dark energy constraints in the absence of any unknown baryonic 
physics.  The results of this exercise represent constraints in the 
limit of perfect knowledge 
of halo profiles and correspond directly to other forecasts in the literature.  
This limit serves as a useful reference for 
interpreting subsequent parameter constraints.  
Following this, we give forecasts for dark energy 
parameter constraints after including and 
marginalizing over the contribution of unknown baryonic physics.  We continue with 
constraints on halo profiles themselves after marginalizing over dark energy.  Our final 
results show dark energy constraints behave as a function of external, 
independent constraints on halo profiles that can be included in the analysis.  
We discuss our results, including caveats and avenues for future study, 
and draw conclusions in Section~\ref{section:conc}.

\section{Methods}
\label{section:methods}

\subsection{Weak Lensing Observables}
\label{subsection:wlo}

In this study, we consider utilizing information in weak lensing observables only.  
We model our approach after that of \citet{ma_etal06}.  In particular, 
we consider convergence power spectra from $\ntomo$ tomographic 
redshift bins defined by the photometric redshifts of source galaxies.  
The $\ntomo (\ntomo+1)/2 $ observables given by the number density-weighted 
convergence spectra and cross spectra for the tomographic bins 
as a function of multipole $\ell$, are 
%
\beq
\label{eq:pkij}
\Pktom{i}{j}(\ell) = \int \dd z  \frac{\wlw{i}(z)\wlw{j}(z)}{H(z)\Da^2(z)}P(k=\ell/\Da,z).
\eeq
In Eq.~(\ref{eq:pkij}), $H(z)$ is the Hubble expansion rate, $\Da$ is the 
angular diameter distance, and $P(k,z)$ is the three-dimensional total matter 
power spectrum at wavenumber $k$ and redshift $z$.  The lensing weight functions 
$\wlw{i}(z)$ specify the tomographic bins.  Defining the true redshift 
distribution of source galaxies in the $i$th photometric redshift bin as 
$\dd n_{\mathrm{i}}/\dd z$, the lensing weight functions are given by 
%
\beq
\label{eq:wi}
\wlw{i}(z) = \frac{3}{2}\Omegam H_0^2 (1+z) \Da(z) 
\int \dd z' \frac{\Da(z,z')}{\Da(z')}\ \frac{\dd n_{\mathrm{i}}}{\dd z'}.
\eeq
The present Hubble rate is $H_0$ and $\Da(z,z')$ denotes 
the angular diameter distance between redshifts $z$ and $z'$.  
Note that the source distribution is not
normalized to integrate to unity and hence $\Pktom{i}{j}$ represents
the power spectra weighted by the product of angular number densities
in the $i$th and $j$th photometric redshift bins.

It is natural to express the photometric redshift bins in terms of 
the total true redshift distribution of source galaxies $\dd n/\dd z$, 
and the probability of yielding a photometric redshift $\zp$ given a 
true redshift $z$, which we denote $P(\zp|z)$.  
In this case, the true redshift distribution of sources 
in the $i$th photometric redshift bin can be obtained by integrating between 
the photometric redshift limits of the bin 
%
\beq
\label{eq:ni}
\frac{\dd n_{\mathrm{i}}(z)}{\dd z}=\int_{z_\mathrm{i}^{\mathrm{low}}}^{z_\mathrm{i}^{\mathrm{high}}} 
\dd \zp \frac{\dd n(z)}{\dd z} P(\zp|z).
\eeq
For the purpose of this study, we take the true redshift distribution to be 
%
\beq
\label{eq:dndz}
\frac{\dd n(z)}{\dd z} = \bar{n}\frac{4z^2}{\sqrt{2\pi}z_0^3}\exp[-(z/z_0)^2]
\eeq
with $z_0 \simeq 0.92$, which fixes the median survey redshift to 
$z_{\mathrm{med}}=1$.  The parameter $\bar{n}$ represents the total density 
of source galaxies per unit of solid angle.  For simplicity and concreteness, we model the 
photometric redshift distribution as in {\tt Model I} of Ref.~\cite{ma_etal06}.  
To be specific, we take $P(\zp|z)$ to be a Gaussian centered at $\zp=z$ with 
dispersion $\sigma_{z}=0.05(1+z)$.  We define tomographic bins to be evenly 
spaced in redshift from $z=0$ to $z=3$, putting whatever fraction (of order $\sim 10^{-4}$) 
of galaxies that lie beyond $z=3$ into the highest redshift tomographic bin.  
We number our tomographic bins sequentially so that $1$ designates the 
lowest redshift bin and $\ntomo$ labels the highest redshift bin.  
Unless we state otherwise, we report results with $\ntomo=5$ as we find 
little relative improvement in parameter constraints 
with finer tomographic binning, in agreement with previous work \cite{ma_etal06}.

\subsection{A Halo Model Primer}
\label{subsection:hm}

To evaluate the relations in Section~\ref{subsection:wlo}, it is necessary to compute the 
nonlinear matter power spectrum.  Though other methods exist and have proven quite successful 
\cite{peacock_dodds96,smith_etal03}, 
we use the phenomenological halo model to accomplish this.  
Elements of the halo model have been developed in a number of studies 
\cite{scherrer_bertschinger91,ma_fry00,seljak00,scoccimarro_etal01,sheth_etal01}.  A comprehensive 
review can be found in Ref.~\cite{cooray_sheth02}.  We implement the halo model as in 
\citet{rudd_etal07} and review the salient features below.

The halo model is predicated on the assumption that all matter resides within dark matter halos.  
The matter power spectrum is given by a sum of two terms, 
%
\beq
\label{eq:psum}
P(k) = \Poneh(k) + \Ptwoh(k).
\eeq
The first term is from matter elements that reside within a common dark matter halo, while the 
second term is due to elements that reside in distinct dark matter halos.  The utility of this 
decomposition in the present context is apparent.  Baryonic processes have a significant 
influence on the structures of individual dark matter halos, but have little effect on the 
large-scale clustering of halos themselves and these ingredients are treated independently 
in the halo model.  

The one-halo contribution is 
%
\beq
\label{eq:p1halo}
\Poneh(k) = \frac{1}{\rhomean^2}\int \dd m\ m^2 \frac{\dd n}{\dd m} \lambda^2(k,m), 
\eeq
where $\rhomean$ is the mean matter density of the universe, $m$ is halo mass, 
$\dd n/\dd m$ is the mass function of dark matter halos, and 
$\lambda(k,m)$ is the Fourier transform of the halo matter density profile, 
which we have assumed to be spherically symmetric and normalized such that 
the integral of the matter density profile $\rho(r,m)$, over all space 
is unity, $4\pi \int \dd r \ r^2 \rho(r,m)=1$.  The two-halo term is 
%
\beq
\label{eq:p2halo}
\Ptwoh(k) = \frac{1}{\rhomean^2}P^{\mathrm{lin}}(k)
\Bigg[\int \dd m\ m \frac{\dd n}{\dd m}\lambda(k,m)\bh(m)\Bigg]^2,
\eeq
where $P^{\mathrm{lin}}(k)$ is the matter power spectrum given by linear 
perturbation theory and $\bh(m)$ is the mass-dependent halo bias.  
We calculate the linear matter power spectrum using the fitting 
formulae of \citet{eisenstein_hu99} with the appropriate modifications 
for dark energy \cite{hu02}.  The halo model 
has known shortcomings (e.g.,~\cite{tinker_etal06}), but we model 
the matter power spectrum in this way because it isolates the quantities 
we aim to study, namely the mean density profiles of halos of mass $m$, 
$\rho(r,m)$ and their Fourier transforms $\lambda(k,m)$.  

Motivated by the findings of Ref.~\cite{rudd_etal07}, 
we model the average mass density of halos 
using the density profile of Ref.~\cite{navarro_etal97}, 
%
\beq
\label{eq:nfw}
\rho(r) \propto \frac{1}{(cr/\Rvir)(1+cr/\Rvir)^2},
\eeq
truncated for $r > \Rvir$.
The parameter $c$ describes the relative concentration of mass toward the halo 
center.  $\Rvir$ gives the extent of halo in the 
sense that the mass within $\Rvir$ is the halo mass $m$.  
We define halos as spherical objects within which the mean density is $200 \rhomean$, 
so that the mass and radius are related by $m=4\pi(200\rhomean)\Rvir^3/3$. This convention 
is chosen to be consistent with the halo bias prescription that we use (see below).  
The density profile of Eq.~(\ref{eq:nfw}) has a Fourier transform \cite{scoccimarro_etal01}
%
\begin{eqnarray}
\label{eq:lam}
\lambda(k,c) & = & \frac{1}{f(c)}\Bigg\{\sin(\eta)[\mathrm{Si}([1+c]\eta)-\mathrm{Si}(\eta)] \nonumber\\
 & & + \cos(\eta)[\mathrm{Ci}([1+c]\eta)-\mathrm{Ci}(\eta)] \nonumber \\
 & & - \frac{\sin(\eta)}{[1+c]\eta}\Bigg\},
\end{eqnarray}
where $f(x)=\ln(1+x)-x/(1+x)$, $\eta=k\Rvir/c$, and $\mathrm{Si}(x)$ and $\mathrm{Ci}(x)$ 
are the sine and cosine integrals respectively.

In the case of dissipationless physics, we set the mean halo concentration as a function of 
mass and redshift according to a power law \cite{bullock_etal01} (see also 
Refs.~\cite{dolag_etal04,wechsler_etal06,maccio_etal07} and in particular Ref.~\cite{neto_etal07} 
for a recent study of concentrations in the 
{\tt Millennium Simulation}~\footnote{{\tt \footnotesize URL http://www.mpa-garching.mpg.de/galform/virgo/millennium/}})
%
\beq
\label{eq:cdm}
c^{\mathrm{STD}}(m,z) = 11[m/m_{\star,0}]^{-0.1}(1+z)^{-1}
\eeq
where $m_{\star,0}$ is the mass of a typical object collapsing at $z=0$ 
($m_{\star,0} \simeq 2.2 \times 10^{12} \hMsun$ in our fiducial cosmology, see 
below).  The slight difference between Eq.~(\ref{eq:cdm}) and the 
relation in Ref.~\cite{bullock_etal01} is due to the conversion between the 
the halo mass definition used in that study and the one we use \cite{hu_kravtsov03}.  
Throughout this study we ignore the distribution in concentrations at 
fixed mass which is approximately log-normal \cite{bullock_etal01}.  
For concentration distributions computed from $N$-body and 
hydrodynamic simulations, the relative influence of the 
spread in concentrations on convergence spectra is 
completely negligible on the scales we consider \cite{cooray_hu01,dolney_etal04}.
For simplicity, we model the halo mass function and the halo bias using the 
relations of Ref.~\cite{sheth_tormen99}.  
We reiterate here that when we refer to halo density profiles and halo 
concentration parameters, we mean the effective density profiles and 
concentrations of the composite systems of dark matter and baryons.

\subsection{Fisher Matrix Analysis, Experimental Specifications, and Cosmological Parameters}
\label{subsection:fm}

We quantify the constraining power of observables 
using the Fisher information matrix.  
The Fisher matrix components are a sum over observables, 
%
\beq
\label{eq:fisher}
F_{\mathrm{ij}}=\sum_{\ell_{\mathrm{min}}}^{\ellmax} (2\ell+1)\fsky 
\sum_{\mathrm{A,B}} \frac{\partial \mathcal{O}_{\mathrm{A}}}{\partial p_{\mathrm{i}}} 
[C^{-1}]_{\mathrm{AB}} 
\frac{\partial \mathcal{O}_{\mathrm{B}}}
{\partial p_{\mathrm{j}}} + F_{\mathrm{ij}}^{\mathrm{P}}, 
\eeq
where the $\mathcal{O}_{\mathrm{A}}$ represent the $\ntomo(\ntomo+1)/2$ observables 
of Eq.~(\ref{eq:pkij}) and the index A runs over these observables, 
$C_{\mathrm{AB}}$ are the components of the covariance matrix of observables, 
and the $p_{\mathrm{i}}$ represent the parameters of the model.  The sum over multipoles 
runs from some minimum multipole $\ell_{\mathrm{min}}$ fixed by the  
sky coverage of the experiment.  For simplicity, we set $\ell_{\mathrm{min}}=2\fsky^{-1/2}$, 
where $\fsky$ is the fraction of the sky covered by the experiment; 
however, this choice is unimportant because constraints are dominated by considerably 
higher multipoles in all cases of interest.  The multipole sum runs to some maximum 
multipole $\ellmax$, which is typically fixed between $\ellmax \sim 1000 - 3000$ 
so as to remain in the regime where several assumptions, including 
that of Gaussian statistics, are valid 
\cite{white_hu00,cooray_hu01,vale_white03,dodelson_etal06,semboloni_etal06}.  
Furthermore, our modeling of baryonic effects is certainly not valid at 
significantly higher multipoles, so it 
is sensible to keep $\ellmax$ in this range for our purposes.  
We elaborate on this in section~\ref{subsection:concentrationmodels}.

The inverse of the Fisher matrix approximates the parameter covariance locally about the 
maximum likelihood.  As such, the measurement error of the $\mathrm{i}$th parameter 
is $\sigma(p_{\mathrm{i}})=[F^{-1}]_{\mathrm{ii}}$.  
The second term in Eq.~(\ref{eq:fisher}) incorporates any Gaussian priors on the model 
parameters.  In most of what follows, we assume modest priors on each parameter 
individually so that $F_{\mathrm{ij}}^{\mathrm{P}}=\delta_{\mathrm{ij}}/\sigma_{\mathrm{i}}^2$, where 
$\delta_{\mathrm{ij}}$ is the Kronecker $\delta$ symbol.  We discuss our parameters 
and priors in more detail below.

The Fisher matrix formalism also provides an estimate of parameter biases due to 
unknown systematic offsets in observables.  Let $\mathcal{O}_{\mathrm{A}}^{\mathrm{bias}}$ be 
the difference between the true observable and the perturbed observable due to some 
systematic effect.  In the limit of small systematic offsets, the parameters extracted 
from this set of observables will be biased by 
%
\beq
\label{eq:bias}
\delta p_{\mathrm{i}} = \sum_{\mathrm{j}} [F^{-1}]_{\mathrm{ij}} 
\sum_{\ell} (2\ell+1) \fsky \sum_{\mathrm{A,B}} 
\mathcal{O}_{\mathrm{A}}^{\mathrm{bias}} [C^{-1}]_{\mathrm{AB}} 
\frac{\partial \mathcal{O}_{\mathrm{B}}}{\partial p_{\mathrm{j}}}.
\eeq

Observed spectra contain both a term due to signal and a noise term, 
%
\beq
\label{eq:pobs}
\Pkobs{i}{j}(\ell)=\Pktom{i}{j}(\ell) +  {n_{\mathrm{i}}} \delta_{\mathrm{ij}} \gi, 
\eeq  
where $\gi$ is the intrinsic source galaxy shape noise, 
and $n_{\mathrm{i}}$ is the surface density of sources in 
the $\mathrm{i}$th tomographic bin and can be obtained by 
integrating Eq.~(\ref{eq:ni}) over all redshifts.  
The covariance between observables $\Pkobs{i}{j}$ and 
$\Pkobs{k}{l}$ is 
%
\beq
\label{eq:cov}
C_{\mathrm{AB}} = \Pkobs{i}{k}\Pkobs{j}{l}+\Pkobs{i}{l}\Pkobs{j}{k},
\eeq
where the $\mathrm{i}$ and $\mathrm{j}$ map 
to the observable index $\mathrm{A}$, 
$\mathrm{k}$ and $\mathrm{l}$ map to $\mathrm{B}$, 
and we have assumed the statistics of the convergence field to be Gaussian.  
We fix $\sqrt{\gi}=0.2$ and study constraints from several forthcoming surveys 
defined by the parameters $\fsky$ and $\bar{n}$, the total effective surface density 
of source galaxies on the sky.  For specificity, we study the limits expected 
from an experiment like the 
Dark Energy Survey (DES)\footnote{{\tt \footnotesize URL http://www.darkenergysurvey.org}} with 
$\fsky=0.12$ and $\bar{n}=15/\mathrm{arcmin}^2$, a future space-based mission like 
the SuperNova/Acceleration Probe (SNAP)\footnote{{\tt \footnotesize URL http://snap.lbl.gov}} with 
$\fsky=0.025$ and $\bar{n}=100/\mathrm{arcmin}^2$, and a future ground-based survey 
like the Large Synoptic Survey Telescope (LSST)\footnote{{\tt \footnotesize URL http://www.lsst.org}}
with $\fsky=0.5$ and $\bar{n}=50/\mathrm{arcmin}^2$.

We consider cosmologies defined by seven parameters.  Three of these 
parameters describe the dark energy.  The other four parameters and 
the values they take in our fiducial cosmological model are 
the total matter density $\omegam=\Omegam h^2=0.128$, 
the baryon density $\omegab=0.0223$, 
the amplitude of curvature fluctuations at $k=0.05 \ \mathrm{Mpc}^{-1}$ 
$\dr=2.04 \times 10^{-9}$ (we actually vary $\ln \dr $), 
and the power-law index of the primordial power spectrum $\ns=0.958$.  
This fiducial model is motivated by the Wilkinson Microwave Anisotropy 
Probe three-year results \cite{spergel_etal07}.  
We take very modest priors on these four parameters, $\sigma(\omegam)=0.007$, 
$\sigma(\omegab)=1.2 \times 10^{-3}$, $\sigma(\ln \dr )=0.1$, and $\sigma(\ns)=0.05$, 
which are comparable to contemporary constraints 
(e.g., Refs.~\cite{tegmark_etal06,spergel_etal07}).  
We describe the dark energy by its present energy density parameter 
$\Omegade=0.76$, and an equation of state parameter that varies with 
scale factor as $w(a)=\wzero+(1-a)\wa$.  In our fiducial model, $\wzero=-1$ and 
$\wa=0$ and we assume no prior information on any dark energy parameter.  
As a reference, our fiducial model implies that the root-mean-square matter 
density fluctuations on a scale of $8 \hMpc$ is $\sigma_8 \simeq 0.76$.

In addition to the marginalized errors in the 
cosmological parameters, we also quantify the constraining power of surveys 
by the figure of merit advocated in the report of the Dark Energy Task Force 
\cite{detf} (see Ref.~\cite{huterer_turner01} for a similar suggestion).  
The DETF figure of merit is the inverse of the area of the 
marginalized 95\% ellipse in the $\wa$-$\wzero$ plane divided by $\pi$.  Letting 
$a$ and $b$ denote the lengths of the principal axes of this ellipse, the figure 
of merit is $\fom = 1/ab$.  Many studies also quote a pivot scale factor $\apiv$, 
and a pivot value of the dark energy equation of state parameter $\wpiv=w(\apiv)$ 
\cite{huterer_turner01,hu_jain04,detf}.  The pivot 
scale factor is the scale factor at which the uncertainty in $w(a)$ is minimized, 
and is given by
%
%
\beq
\apiv = 1 + \frac{[F^{-1}]_{\wzero \wa}}{[F^{-1}]_{\wa \wa}}.
\eeq
The pivot equation of state parameter is simply 
%
\beq
\wpiv = \wzero + (1-\apiv)\wa.
\eeq
Taking $\wa$ along with $\wpiv$ as the parameters describing the dark energy equation of 
state is convenient because the principal axes of the error ellipse lie along these 
coordinates.  Moreover, the transformation is linear and preserves the area of the
ellipse.  In terms of the components of the Fisher matrix [Eq.~(\ref{eq:fisher})], 
the variance in $\wpiv$ is given by 
%
%
\beq
\sigma^2(\wpiv) = [F^{-1}]_{\wzero \wzero} - \frac{[F^{-1}]^2_{\wzero\wa}}{[F^{-1}]_{\wa\wa}}.
\eeq
The 95\% confidence level in two dimensions occurs at $\simeq 2.48\sigma$, so that 
the area of the 95\% ellipse is ${\cal A} \simeq 6.17\pi\sigma(\wa)\sigma(\wpiv)$.  
The DETF also quote values of $[\sigma(\wa)\sigma(\wpiv)]^{-1}$ in their summary 
tables \cite{detf}.  Clearly, our figure of merit $\fom$, 
is related to the numerical values quoted in the tables of the  DETF 
report $[\sigma(\wpiv)\sigma(\wa)]^{-1}$, by
%
%
\beq
\fom \simeq \frac{0.162}{\sigma(\wpiv)\sigma(\wa)}.
\eeq
We intend to focus on the relative scaling of the figure of merit, so the 
details of the constants of proportionality here are of minimal importance.

\subsection{Modeling the Effects of Baryons with Modified Halo Concentrations}
\label{subsection:concentrationmodels}

\citet{rudd_etal07} demonstrated that the influence of baryons on the matter power spectrum 
is primarily due to the modified structure of individual dark matter halos.  In fact, 
the modification is simple.  \citet{rudd_etal07} found that the halos in their hydrodynamic 
simulations followed a concentration-mass relation that was significantly different from 
the relation derived from $N$-body simulations [e.g., Eq.~(\ref{eq:cdm})].  \citet{rudd_etal07} 
found that the description of halo structure as NFW halos with boosted concentrations was valid 
for halo-centric radii greater than approximately $0.04\Rvir$
\footnote{The difference in numerical factors reported here reflects the difference between the halo 
definition used in Ref.~\cite{rudd_etal07} and the one we use in this study.  We have converted 
between the two definitions as described in Ref.~\cite{hu_kravtsov03}.}. 
Upon accounting for this difference, these authors found that they could model convergence 
power spectra derived from their simulations accurately out to multipoles as high as 
$\ell \sim 6000$.  However, as noted in both Ref.~\cite{jing_etal06} and Ref.~\cite{rudd_etal07}, 
hydrodynamic simulations are not yet up to the task of providing accurate predictions for the 
properties of galaxies.  So, while the net effect of galaxy formation on the power spectra may be 
well described by a modified, effective $c(m,z)$ relation, this effective $c(m,z)$ relation 
is not known.  Consequently, it seems natural to calibrate the effective 
concentration-mass relation within the observational data itself 
and we explore several parameterizations of varying complexity for this relation.
In the following section, we present results where both cosmology and the parameters 
of the $c(m,z)$ relation are determined simultaneously from observational data.

Neglecting baryonic physics, the relation between concentration and mass is known to be 
well characterized by a power law [Eq.~(\ref{eq:cdm}), see Ref.~\cite{bullock_etal01,neto_etal07}].  
Consequently, it seems natural to study power-law relations of the form 
%
\beq
\label{eq:plc}
c(m,z) = c_0 [m/m_{\star,0}]^{\alpha}(1+z)^{-\beta}.
\eeq
Further, it is reasonable to suppose that the redshift dependence is not modified from 
the standard $N$-body relation [Eq.~(\ref{eq:cdm})] because the majority of baryonic 
condensation is thought to have occurred at $z \gsim 1$ and because the redshift scaling in 
the standard case arises simply because virial radii grow as $\Rvir \propto (1+z)^{-1}$ 
in the absence of accretion.  This scaling arises because halos are typically defined 
as an overdensity with respect to the mean density and the mean density dilutes as 
$ \rhomean ~ (1+z)^{-3}$ while the structure of the bound object remains unchanged.  
We expect this second feature to be unmodified when including 
baryonic physics.  Nevertheless, we study concentration relations of this form both with 
the redshift dependence fixed to $\beta=1$ and with $\beta$ free.  We choose a fiducial 
concentration relation that is enhanced so that $c_0=15$ and $\alpha=-0.1$. 
This represents a $36\%$ enhancement in halo concentrations while 
\citet{rudd_etal07} report a slightly larger enhancement in concentrations in their 
{\tt DMG\_SF} simulation for halos near $m ~ \mathrm{a\ few} \ \times 10^{14} \hMsun$.  
As we demonstrate in Section~\ref{section:results}, 
it is the halos in this mass range that most influence convergence power spectra.  
We choose an enhancement slightly below the most relevant result from Ref.~\cite{rudd_etal07} 
because the simulation results likely over-estimate the effective concentrations 
due to the well-documented overcooling problem of contemporary hydrodynamic simulations 
\cite{katz_white93,suginohara_ostriker98,lewis_etal00,pearce_etal00,dave_etal01,balogh_etal01,borgani_etal02}.

Though concentration relations normalized at $m=m_{\star,0}$ and redshift $z=0$ are 
conventional in the literature on halo structure, this is likely not to be the most appropriate  
representation in the present context.  Halos near $m \sim 10^{14} \hMsun$ make the largest 
contribution to convergence power spectra \cite{CooHuMir00,zhan_knox04}, and lensing weights are considerable at 
redshifts between $ 0.2 \lesssim z \lesssim 1$, so it is natural to suppose that lensing will be most 
sensitive to the structure of $\sim 10^{14} \hMsun$ halos in a broad range of redshifts.  
In analogy with the the pivot scale factor and pivot equation of state 
parameter described in Section~\ref{subsection:fm} for the dark 
energy, we may define a pivot redshift $\zpiv$, and pivot mass $\mpiv$, for the $c(m,z)$ relation.  
We set the $\zpiv$ and $\mpiv$ as the redshift and 
mass at which the fractional uncertainty $\sigma(c)/c$ is minimized.  
The pivot concentration is $\cpiv = c(\mpiv,\zpiv)$.  
The fractional uncertainty at the pivot concentration is 
%
%
\begin{eqnarray}
\label{eq:cpiv}
\frac{\sigma^2(\cpiv)}{\cpiv^2} & = & 
\frac{[F^{-1}]_{c_0 c_0}}{c_0^2} + \ln^2(\mpiv/m_{\star,0})[F^{-1}]_{\alpha \alpha} \nonumber \\
 & + & \ln^2(1+\zpiv) [F^{-1}]_{\beta \beta} \nonumber \\
 & + & \frac{2 \ln(\mpiv/m_{\star,0})}{c_0} [F^{-1}]_{c_0 \alpha} \nonumber \\
 & - & \frac{2 \ln(1+\zpiv)}{c_0} [F^{-1}]_{c_0 \beta} \nonumber \\
 & - & 2 \ln (\mpiv/m_{\star,0}) \ln(1+\zpiv) [F^{-1}]_{\alpha \beta}.
\end{eqnarray}
The pivot redshift is 
%
%
\beq
\label{eq:zpiv}
\ln (1+\zpiv) = \frac{1}{c_0} 
\frac{[F^{-1}]_{\alpha \alpha} [F^{-1}]_{c_0 \beta} - [F^{-1}]_{c_0 \alpha} [F^{-1}]_{\alpha \beta}}
{[F^{-1}]_{\alpha \alpha} [F^{-1}]_{\beta \beta} - [F^{-1}]_{\alpha \beta}^{2}}
\eeq
and the pivot mass is 
%
%
\beq
\label{eq:mpiv}
\ln (\mpiv/m_{\star,0}) = \frac{1}{c_0}
\frac{[F^{-1}]_{\beta \beta} [F^{-1}]_{c_0 \alpha} - [F^{-1}]_{c_0 \beta} [F^{-1}]_{\alpha \beta}}
{[F^{-1}]_{\alpha \beta}^{2} - [F^{-1}]_{\alpha \alpha} [F^{-1}]_{\beta \beta}}.
\eeq
In the following sections, we will quote pivot masses and redshifts and errors in 
$\cpiv$.

The $c(m,z)$ relation of Eq.~(\ref{eq:plc}) may be general enough to encompass most 
plausible behavior, particularly because the convergence spectra are most sensitive 
to a narrow range of halo masses near $m \sim 10^{14} \hMsun$ for
$\ell \sim 10^{3}$ \cite{CooHuMir00,zhan_knox04} (also see Sec.~\ref{section:results}).  
However, it is not unreasonable to suspect more complicated behavior.  For 
example, the mass-to-light ratios of galaxy, group, and cluster halos are 
known to vary in a non-monotonic fashion with total mass 
(e.g., Ref.~\cite{bahcall_etal00,tinker_etal05,vdb_etal07}) with a 
minimum near $m \sim 10^{11.5} \hMsun$.  As this minimum reflects some maximum 
efficiency for galaxy formation, it is natural to suspect that the relative 
boost in halo concentrations peaks at this halo mass and declines on either 
side of it.  To hedge against the possibility of more complicated 
behavior than that in Eq.~(\ref{eq:plc}), we also explore a concentration relation
with significantly more freedom.

In this more general relation we define values of concentration
at $N_{\mathrm{c}}$ values of $\log(m/\hMsun)$ evenly spaced in the log of 
the halo mass between $11 \le \log(m/\hMsun) \le 16$.  
We have verified that extending this range 
to lower or higher masses has little effect as these masses 
contribute too little to the convergence spectra on scales of interest \cite{CooHuMir00,zhan_knox04}.  
The concentration values specify the mean halo concentration at the center of each
mass bin and we spline interpolate between these values to enforce smoothness in 
the relation.  This choice reflects our prejudice that the average concentration
should always be a very smooth function of mass.  
In addition, we expect that at very high masses and very low masses 
baryonic condensation and galaxy formation should be inefficient, so we require 
that the standard relation [Eq.~(\ref{eq:cdm})] be recovered for masses 
$\log(m/\hMsun) < 9$ and $\log(m/\hMsun) > 18$.  
We enforce this by matching the spline onto the standard relation at 
these endpoints.  In practice, the choice of boundary condition for the 
$c(m,z)$ at low and high masses is unimportant.  We refer to our parameter set by 
subscripting a ``$c$'' with the logarithm of the mass at which this concentration 
is specified.  For example, in a model with $N_{\mathrm{c}}=5$, our concentration parameters 
are $c_{11.5}$, $c_{12.5}$, $c_{13.5}$, $c_{14.5}$, and $c_{15.5}$.  
We implement a variable redshift dependence as $c(m,z)=c(m,z=0)/(1+z)^{\beta}$.  

\begin{figure}[t]
\begin{center}
\includegraphics[height=12cm]{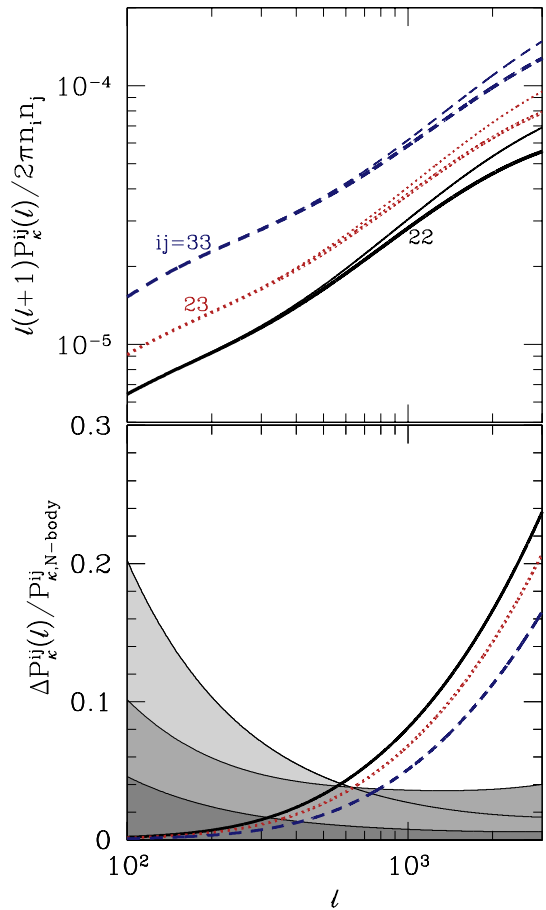}
\caption{
The differences in convergence spectra for the fiducial model of boosted power-law 
concentrations ({\em thin lines}) relative to the standard case with concentrations 
derived from dissipationless simulations ({\em thick lines}).  In the top panel we show the 
power spectra explicitly and in the bottom panel we show the change in observables 
scaled to the standard $N$-body concentration case.  We use five tomographic 
bins throughout most of this work and we have dropped the number weighting of 
spectra for the purposes of this demonstration only.  
For simplicity we show only the behavior 
of three observables, the spectra in the second and third redshift bins 
$\Pktom{2}{2}/n_{2}^2$ and $\Pktom{3}{3}/n_{3}^2$, 
and the cross spectrum $\Pktom{2}{3}/n_{2}n_{3}$.  
The behavior of other observables is qualitatively similar.  The shaded 
bands show, from outermost to the innermost at left, the variance in $\Pktom{3}{3}$ 
for SNAP, DES, and LSST in bands of width $\Delta \ell/\ell = 1/10$.  Note that 
the sensitivity of SNAP overtakes that of DES by $\ell \sim 600$ as SNAP is far 
deeper than DES.  We show this by extending the upper limit of the shaded SNAP 
band as  a solid line below the filled DES region.  
}
\label{fig:pkcomp}
\end{center}
\end{figure}

To give the reader a sense of the influence of these changes on observables, 
we show the relative differences in the observable convergence spectra in 
Figure~\ref{fig:pkcomp}.  For concreteness, we plot spectra in the power 
law $c(m,z)$ model with $c_0=15$ compared to the spectra computed using the 
standard relation of Eq.~(\ref{eq:cdm}) derived from dissipationless simulations.  
We have used a model with five tomographic redshift bins, but for the 
sake of clarity we plot only three observables.  These correspond to the spectra of the second and 
third tomographic bins and the cross spectrum between these bins.  As a reminder, 
the {\em photometric} redshift limits of the second and third bins are 
$0.6 < \zp \le 1.2$ and $1.2 < \zp \le 1.8$ respectively.  As one would expect, 
any effect on concentrations becomes quite large at $\ell \gsim 10^{3}$.  However, 
it is worth noting that the effect is not negligible even at much lower multipoles 
and that all of the experiments we consider would be sensitive to such effects 
even at multipoles as low as several hundred.

\section{Results}
\label{section:results}

\subsection{Dark Energy Constraints and Parameter Biases}
\label{subsection:destandard}

We begin our presentation of results with parameter constraints in the standard case of 
no modifications to halo structure from baryonic processes.  This case corresponds 
to perfect knowledge of halo density profiles and has been  
considered numerous times in the literature already.  
We include this partly to give a frame of 
reference for the results that follow.  More importantly, we show 
dark energy parameter constraints as a function of the maximum multipole $\ellmax$ 
used in the sum in Eq.~(\ref{eq:fisher}) so that one can compare the relative merits 
of marginalizing over an unknown halo concentration-mass relation with simply ignoring 
the high multipole moments where baryonic effects are most severe.  Throughout this 
section, we label the uncertainties in parameter $p$ 
obtained by including multipoles out to $\ellmax=\ell$ 
as $\sigma_{\ell}(p)$.

\begin{figure}[t]
\begin{center}
\includegraphics[height=10cm]{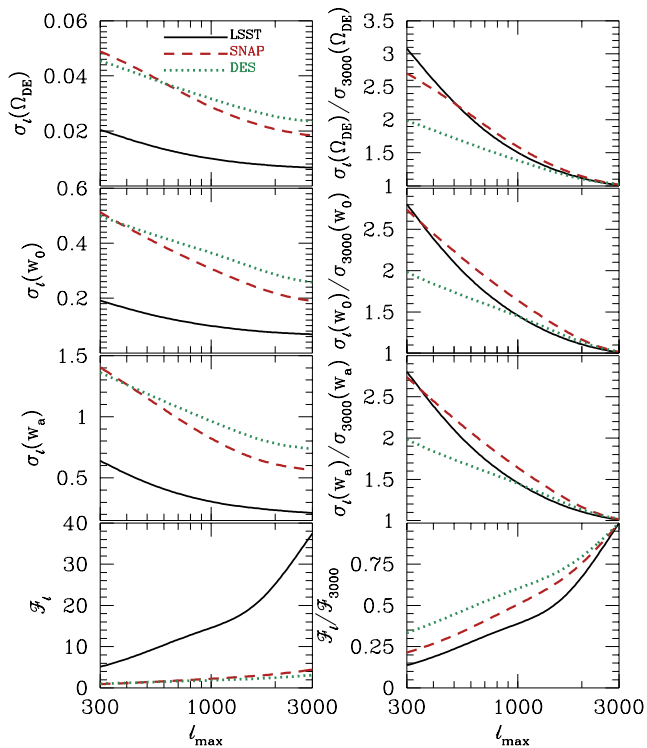}
\caption{
Standard marginalized dark energy parameter constraints as a function of 
maximum multipole $\ellmax$.  The left panels show the $1\sigma$ uncertainties 
in the three dark energy parameters $\Omegade$, $\wzero$, and $\wa$ from the top down.  
The bottom panel shows the dark energy figure of merit $\fom$ as a function of maximum 
multipole.  The {\em solid} line corresponds to an LSST-like experiment, 
the {\em dashed} line corresponds to a SNAP-like experiment, and the {\em dotted} line 
corresponds to an experiment similar to DES.  The right panels show the same quantities 
scaled by the uncertainties obtained at $\ellmax=3000$ in order to show the relative 
scaling of dark energy constraints with $\ellmax$ more clearly.
}
\label{fig:sigma_std}
\end{center}
\end{figure}

Constraints on dark energy parameters in the absence of any unknown baryonic processes 
as a function of $\ellmax$ are shown in Figure~\ref{fig:sigma_std}, where we have taken 
$\ntomo=5$ as prescribed by Ref.~\cite{ma_etal06}.  Unless otherwise noted, this choice 
pertains to all results on parameter constraints.  
With modest prior information, weak lensing tomography alone can constrain 
dark energy parameters very well.  Using information out to an $\ellmax=3000$, 
the marginalized constraints from DES, SNAP, and LSST respectively are 
$\sigma_{3000}^{\mathrm{STD}}(\Omegade)=0.023,0.018,0.0066$, 
$\sigma_{3000}^{\mathrm{STD}}(\wzero)=0.26,0.19,0.069$, 
and $\sigma_{3000}^{\mathrm{STD}}(\wa)=0.74,0.56,0.21$.  
Constraints from LSST are the most stringent because the experiment combines 
extensive sky coverage with depth. SNAP and DES are comparable 
(considering {\em only} weak lensing information) over this range of scales.  
SNAP, being extraordinarily deep, can exploit information in very high multipoles while 
the constraints from DES overtake those from SNAP if only multipoles $\ell \lsim 400$ 
are used because of its larger sky coverage.  
Hereafter, we plot our results including 
self-calibration of baryonic processes relative to these standard results which represent 
the limit of perfect knowledge of internal halo structure.

Both \citet{jing_etal06} and \citet{rudd_etal07} pointed 
out that the modifications of the 
convergence power spectra due to baryonic physics are 
likely to be large and to extend over 
a wide range of multipoles, but neither of these studies 
translated this systematic into 
its influence on cosmological parameter constraints.  
In particular, it is interesting to 
examine the bias in dark energy parameters if baryonic
physics are neglected.  If the bias is small 
compared to the statistical uncertainties in Fig.~\ref{fig:sigma_std}, 
then it need not be a major concern because it is unlikely to lead to rejection 
of the true cosmology based on these data.

\begin{figure}[t]
\begin{center}
\includegraphics[height=9cm]{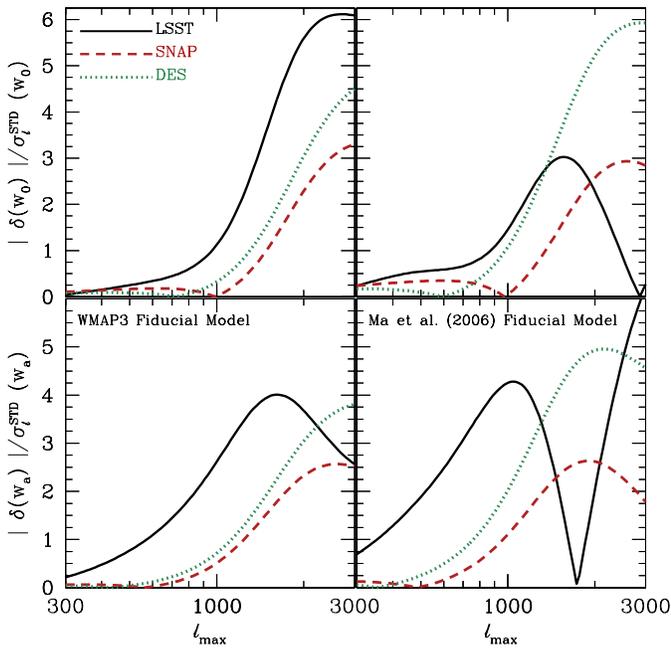}
\caption{
Biases in inferred dark energy parameters due the ignorance of baryonic physics.  
We show biases by assuming the true model to have a power-law concentration-mass 
relation with $c_0=15$ instead of $c_0=11$ as in the case of models that neglect 
baryonic physics.  Results for DES, SNAP, and LSST are shown as labeled in the upper 
left portion of the figure.  In all cases, we show bias relative 
to the 1$\sigma$ measurement uncertainty in the relevant parameter.  
The top panels show the biases for $\wzero$ and the bottom panels show 
the biases for $\wa$.  Notice that we show biases about both the 
fiducial model that we use in this work ({\em left column}) 
and the model of \citet{ma_etal06} which has significantly 
more small-scale power ($\sigma_8 \simeq 0.9$, {\em right column}).  
The minima in the magnitudes of the biases reflect changes in sign.  
Generally $\delta(\wzero) > 0$ at the lowest multipoles and 
$\delta(\wa) < 0$ at the lowest multipoles.  In the case of LSST, 
the first change in sign in $\delta(\wzero)$ occurs at $\ellmax < 300$, 
so that $\delta(\wzero) < 0$ at $\ellmax=300$.  
The approximations used to compute the parameter bias 
[see Eq.~(\ref{eq:bias})] break down in all cases as 
the biases become larger than their uncertainties at high multipoles.  
These estimates indicate that the biases are likely to be non-negligible.
}
\label{fig:biases}
\end{center}
\end{figure}

We show the biases in the inferred values of $\wzero$ and $\wa$ 
as a function of $\ellmax$ in units of their uncertainties in 
Figure~\ref{fig:biases}.  To compute the biases shown in 
Fig.~\ref{fig:biases}, we assumed the true power spectra 
to be given by the fiducial power-law concentration model 
with $c_0=15$, which is a $\sim 36\%$ enhancement 
relative to the standard concentration-mass 
relation, and employed the approximation of Eq.~(\ref{eq:bias}).  
The magnitudes of the 
biases show sharp minima in many cases.  These represent changes in 
the signs of the biases.  Generally, 
$\delta(\wzero) > 0$ and $\delta(\wa) < 0$ at the lowest multipoles.  
Note that in all cases biases become large compared to the 
uncertainties in the parameters, so 
the approximation used to compute them breaks down.  
These large biases are unlikely to be the product of any 
analysis.  Rather the team performing the analysis would 
notice something amiss because the best-fitting models 
would be relatively poor fits to the data and this would be 
evident in some criterion used to assess fit quality, 
such as the $\chi^2$ per degree of freedom.

The details of the parameter biases are a strong 
function of the choice of fiducial cosmology.  This is because 
both the signal-to-noise ratios and the scales at which nonlinear effects 
become important are strong functions of the fiducial model.  We 
illustrate this in Fig.~\ref{fig:biases} by showing biases computed 
about the fiducial cosmological model of \citet{ma_etal06}, which more 
closely resembles the cosmology from the Wilkinson Microwave Anisotropy 
Probe first year results \cite{spergel_etal03}.  The most 
important differences between our fiducial model and the fiducial model of \citet{ma_etal06} 
are that \citet{ma_etal06} took $\ns=1$ (compared to $\ns=0.958$) and a 
slightly higher power spectrum normalization, implying a larger mass 
variance on $8 \hMpc$ scales of $\sigma_8 \simeq 0.9$.  
In the \citet{ma_etal06} fiducial model, parameter 
inferences are more sensitive to small systematics on quasi-linear and 
nonlinear scales for two reasons.  First, the signal is relatively higher.  
Second, nonlinear effects become important on larger scales (smaller $\ell$) 
in models with more power.  Both of these effects are 
reflected in Fig.~\ref{fig:biases}.  Near the \citet{ma_etal06} fiducial model 
with relatively greater power, biases are typically larger and they become 
important at smaller wavenumbers.  Throughout the remainder of this section, 
we give our primary results derived about the fiducial model of Section~\ref{subsection:fm}; 
however, we often compare these to results derived about the fiducial model 
of \citet{ma_etal06} in order to indicate the importance of the choice of 
fiducial model.

As the details of the fiducial model have a significant influence on 
parameter biases and the linearized relation of Eq.~(\ref{eq:bias}) 
should break down for biases large compared to their uncertainties, 
it is hard to make concrete statements about parameter biases.  
Nevertheless, Fig.~\ref{fig:biases} is an indicator that 
such biases may be significant, even at multipoles $\ell < 10^3$.  
LSST will provide the 
tightest constraints on dark energy parameters over this range of 
scales, primarily because it is extremely wide, 
yet this makes LSST most susceptible 
to unknown baryonic physics in a relative sense.  
In the absence of reliable and robust predictions for power spectra in 
the presence of baryons or 
any other method for dealing with the influence of baryonic processes, one could safely 
force the biases to be relatively small compared 
to the uncertainties by disregarding multipoles above some $\ellmax$.  This and 
more elaborate strategies for removing some of the small-scale information in weak lensing 
tomography have been studied by \citet{huterer_white05}.  Fig.~\ref{fig:biases}
suggests that this would require setting $\ellmax$ to no greater than many hundreds, 
though the details depend upon both the experiment (they are most stringent for LSST) 
and the underlying cosmology.  
Comparing with Fig.~\ref{fig:sigma_std}, this corresponds to a 
degradation of dark energy parameter constraints 
by a factor of $\sim 2-3$ or more relative to the constraints achieved 
by setting $\ellmax=3000$ in the 
standard case of perfect knowledge of halo density profiles.  
Correspondingly, the decrease in the DETF figure of merit is 
approximately a factor of $\sim 4-7$.  
We note that these effects are more important in models with more 
small-scale power, so that about a cosmology with 
a higher value of $\sigma_8 \simeq 0.9$, dark energy parameter 
constraints can be degraded by more than a factor of three and 
the figure of merit can be decreased by nearly a factor of ten.  
In what follows, we show that internal self calibration of 
baryonic effects may be a viable alternative to this degradation 
in parameter constraints.

\subsection{Dark Energy Constraint Degradation:  Power-Law Halo Mass-Concentration Relations}
\label{subsection:plc}

\begin{figure}[t]
\begin{center}
\includegraphics[height=10cm]{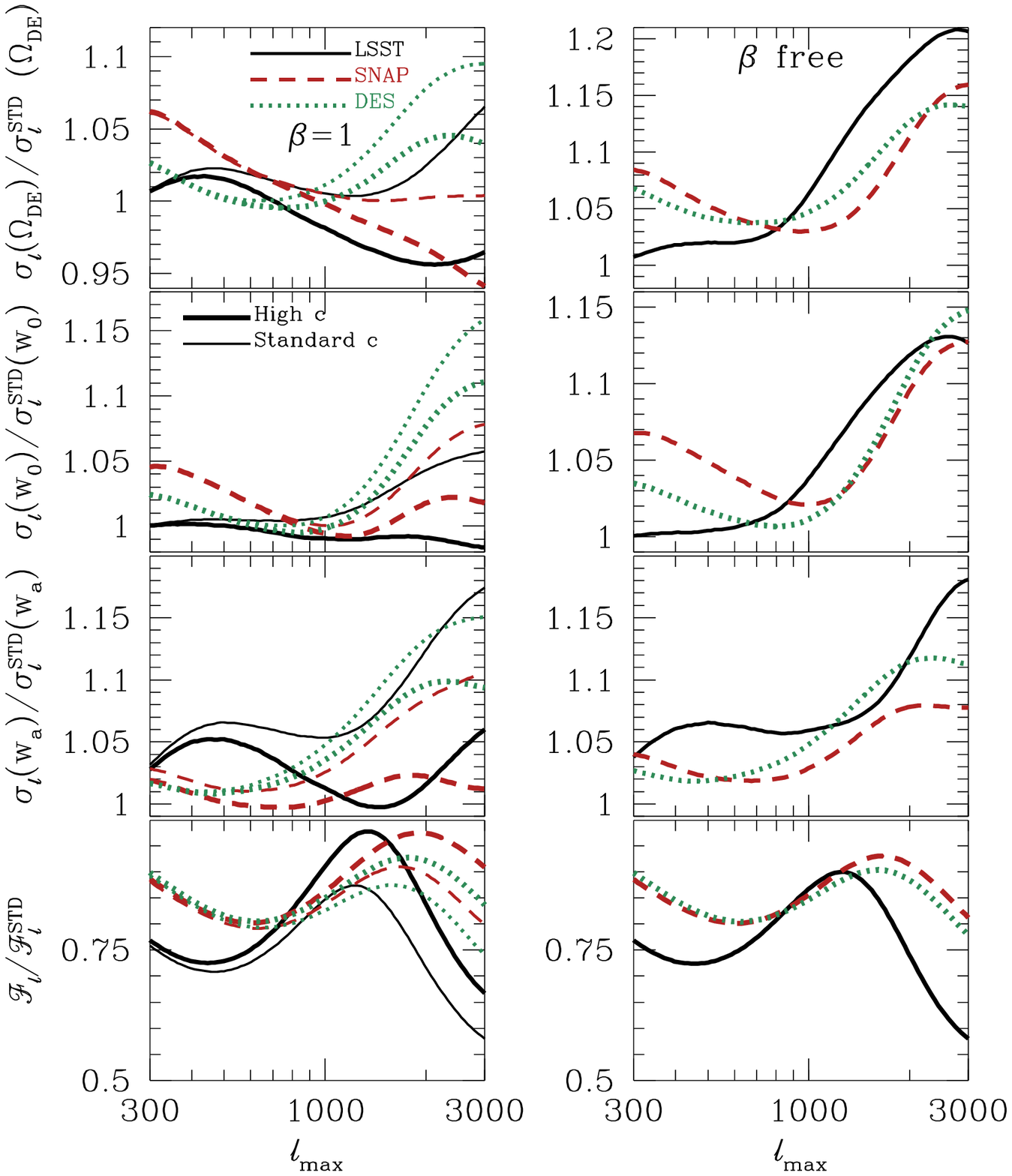}
\caption{
The relative degradation in parameter constraints as a function of maximum multipole 
used to derive constraints for the case of a power-law concentration 
mass relation [Eq.~(\ref{eq:plc})].  The vertical axes are the parameter constraints 
in the new concentration model divided by the constraints in the standard case.  In the 
{\em left} panel, we show constraints in a model with fixed redshift dependence.  The 
{\em thick} lines show constraints about a fiducial model with concentrations enhanced 
relative to the standard $N$-body case ($c_0=15$) while the 
{\em thin} lines show constraints about the $N$-body $c^{\mathrm{STD}}(m,z)$ ($c_0=11$) as fiducial 
model [Eq.~(\ref{eq:cdm})].  In the {\em right} panel we show constraints in a model in 
which the power-law index describing the redshift dependence of concentrations is variable.  
The fiducial model in this case and in all subsequent plots has enhanced halo 
concentrations relative to the dissipationless case.  
Note that the dynamic ranges on the vertical axes are different in the left and 
right panels.
In all panels, the {\em solid} line represents LSST, the {\em dashed} line 
represents SNAP, and the {\em dotted} line represents DES.
}
\label{fig:selfcalpl}
\end{center}
\end{figure}

We consider the degradation in parameter constraints sequentially from the power-law $c(m,z)$ model 
to the more general cases in order to give a sense of the gradual degradation from the additional 
degrees of freedom.  The relative degradation in dark energy parameters in the power-law 
$c(m,z)$ models is shown in Figure~\ref{fig:selfcalpl}.  The left panel of Fig.~\ref{fig:selfcalpl} 
shows constraints with a fixed redshift dependence and the right panel shows constraints with a 
variable redshift dependence parameter $\beta$.  This figure shows that the degradation of 
dark energy parameter constraints is relatively small and that it behooves one to model 
the unknown baryonic effects and continue to use high multipole moments rather than cut out 
information above some relatively low $\ellmax$.  Provided there is some compelling justification 
for these relatively restricted concentration models, we can eliminate the bias caused by 
unknown halo structure while dark energy constraints will be degraded by 
only $\sim 15-20\%$ or less in all cases.  This cost is relatively low compared to 
the cost of using only multipoles below several hundred to constrain dark energy and, 
as we discuss below, the nuisance parameters of this analysis are physical parameters 
of interest in their own right.

Aside from this overriding point, there are several other features worthy of note.  
First, notice that there are instances (depending upon multipole range and experiment) in which 
the model with extra freedom provides better constraints than the standard model with known 
halo concentrations.  This is due to our assumption that concentrations are enhanced 
(we take $c_0=15$ in our model with baryons compared to $c_0=11$ as given by $N$-body simulations)  
resulting in increased signal over a wide range of scales (See Fig.~\ref{fig:pkcomp}).  
The thin lines in 
the left panel of Fig.~\ref{fig:selfcalpl} show constraints about the standard $c(m,z)$ 
relation with $c_0=11$ as fiducial model.  In this case, there is 
no boost in signal and the constraints 
are always poorer than in the model with perfect knowledge of $c(m,z)$, though not by much.  
Hereafter, we quote results for a fiducial model with enhanced halo concentrations 
relative to the standard $N$-body values.  In all cases, the choice of fiducial 
cosmological parameters affects parameter constraints 
at levels similar to those shown in the left panel of 
Fig.~\ref{fig:selfcalpl}.

In addition, observe that in the panels of Fig.~\ref{fig:selfcalpl} the relative degradation 
has a peak at multipoles of a few thousand with some variation depending upon the parameter of 
interest and the experiment.  The peaks are more prominent in the right panels.  
In fact these features would also be apparent in the left panels of 
the figure, but are mostly omitted as they lie beyond our choice of $\ellmax=3000$.  The 
origins of these features are simple to understand.  
The effect of concentrations is predominantly a 
high-$\ell$ phenomenon.  As one approaches $\ell \sim 2000$ from the low multipole side, the 
unknown concentrations are degrading dark energy constraints rapidly because their effects 
are just beginning to appear and can at some level be mimicked by the dark energy parameters.  
However, as one continues to higher multipoles ($\ell \gsim 3 \times 10^{3}$), 
the concentration effects become increasingly distinguishable from those 
attributable to dark energy parameters, these degeneracies are broken and constraints 
improve to higher multipoles thereafter.  The low-$\ell$ feature in these plots is due to a similar 
degeneracy primarily between the power-law index $\alpha$ and the dark energy 
equation of state parameters.  
Such features caused by the gradual elimination of degeneracies 
with the inclusion of information from ever higher multipoles are generic in the concentration 
models that we explore and are present in subsequent plots as well.  
In all cases, these ``rolling hills'' represent the sequential introduction and 
elimination of parameter degeneracies and we do not expound upon the details of each 
degenerate direction in this study in the interest of brevity.

\begin{figure}[t]
\begin{center}
\includegraphics[height=7cm]{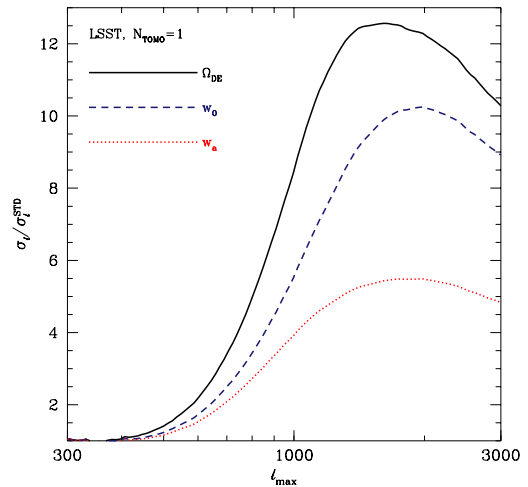}
\caption{
The degradation in dark energy parameter constraints without tomographic 
information.  This figure shows the relative errors compared to the case of 
perfect knowledge of effective halo concentrations as a function of 
$\ellmax$ as in Fig.~\ref{fig:selfcalpl}, but with no tomography.  To illustrate 
this point, we use the power law $c(m,z)$ model with variable redshift index 
$\beta$.  For the sake of clarity, 
we show only the case of LSST here, but the 
loss of constraining power is dramatic for all experiments.  
}
\label{fig:notom}
\end{center}
\end{figure}

It is important to stress that tomographic information is a crucial ingredient in 
minimizing parameter constraint degradation in the case of unknown halo concentrations.  
For example, eliminating tomographic information (setting $\ntomo=1$) dramatically reduces 
the ability of such surveys to calibrate for unknown halo concentrations. 
We find that in the case of the power-law class of concentration models with variable 
redshift index $\beta$, eliminating tomographic information 
in both the concentration-marginalized results and the $N$-body standard 
causes the relative degradation to rise from the $\sim 15\%$ level as shown 
in Fig.~\ref{fig:selfcalpl} to a factor of $\sim 5$ for $\wa$ and a factor of $\sim 10$ 
for $\Omegade$ and $\wzero$.  Note that this degradation is with respect 
to an already degraded baseline with no tomographic information.  
Figure~\ref{fig:notom} shows an example of the 
degradation in dark energy parameter constraints without tomography 
for the LSST experiment.  Moreover, note that this level of dark energy information loss 
extends to multipoles as low as $\ellmax \sim 600$.  On the other hand,
 we find that constraints on dark energy parameters effectively 
saturate at $\ntomo=5$ getting only very gradually better with increasingly fine 
photometric redshift binning.  For instance, increasing to $\ntomo=10$ only gives a 
few percent decrease in dark energy parameter errors.

The importance of tomographic information has both positive and negative aspects.  
On the negative side, it stresses the need for accurate photometric redshifts and 
photometric redshift distributions with well-understood properties.  However, it 
 suggests that the ability of future surveys to self calibrate does not rely 
solely on the shapes of the power spectra produced by modified baryonic physics 
being non-degenerate with those arising due to dark energy 
effects.  To be more specific, this also suggests that this success is not 
limited to very restrictive $c(m,z)$ prescriptions.  
For example, if the power spectra shapes obtained by 
modifying halo concentrations were in no way degenerate with the shapes 
obtained by varying dark energy parameters, eliminating tomographic 
information would not lead to further degradation in parameter constraints because 
even in this case the non-degenerate parameters could not compensate for each other.  
That constraints do significantly degrade in the case of no tomography suggests that 
dark energy and concentrations may lead to degenerate power spectra but that 
tomography helps distinguish the two.  Tomographic information effectively provides several 
handles on the relative distances to the sources in the different photometric 
redshift bins [Eq.~(\ref{eq:wi})] and enables one to distinguish between 
contributions to the shear from high redshift and low wavenumber from 
contributions at high wavenumber and low redshift, both of which may 
contribute at a single multipole.  These internal consistency 
checks within weak lensing tomography render the dark energy and concentration 
parameters relatively distinguishable.  Already, this suggests that the more 
general concentration relations we describe below will not lead to drastic 
dark energy parameter degradation.

\subsection{Constraint Degradation:  Binned Halo Mass-Concentration Relation}
\label{subsection:binnedc}


\begin{figure}[t]
\begin{center}
\includegraphics[height=10cm]{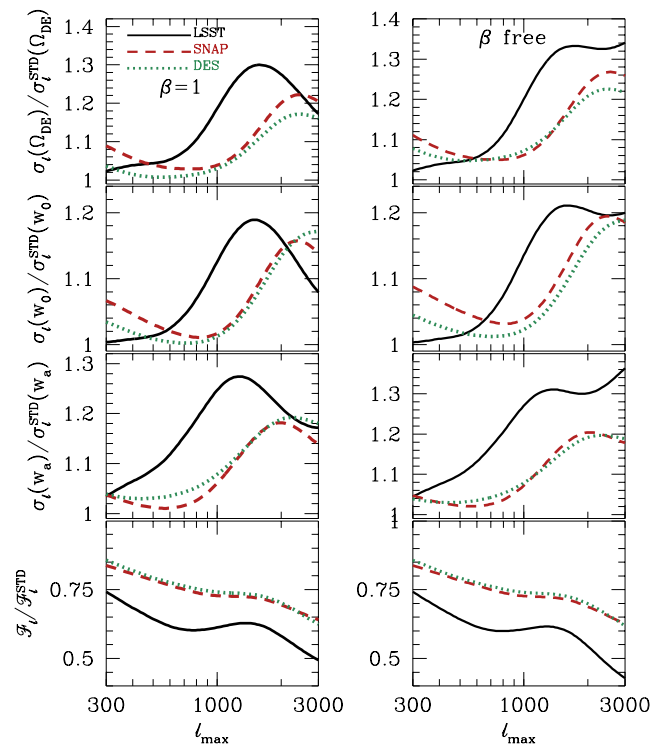}
\caption{
The relative degradation in parameter constraints as a function of maximum multipole 
used to derive constraints for the case of the binned concentration 
mass relation.  The vertical axes are the parameter constraints 
in the new concentration model divided by the constraints in the standard case.  In the 
{\em left} panel, we show constraints in a model with fixed redshift dependence.  
In the {\em right} panel we show constraints in a model in 
which the power-law index describing concentration redshift dependence is variable.  
In all panels, the {\em solid} line represents LSST, the {\em dashed} line 
represents SNAP, and the {\em dotted} line represents DES.
}
\label{fig:selfcal5bin}
\end{center}
\end{figure}

We now present parameter constraints in the context of our more general, 
binned $c(m,z)$ models.  Based on the discussion of the previous section, 
there is cause to be optimistic that self calibration with a more general 
relation for halo concentrations will quite successful.  Indeed, 
Figure~\ref{fig:selfcal5bin} reveals this to be the case.  Maximal parameter 
degradation is $\sim 30\%$ or less in all cases, so again self calibration 
performs better than the more drastic measure of disregarding the information 
contained in multipoles larger than a few hundred.  
Figure~\ref{fig:selfcal5bin} depicts the case of a 
five-bin concentration-mass relation for concreteness; however, 
we have found that the level of degradation is relatively robust to the 
number of bins used.  In fact, the level of degradation remains 
below a worst case degradation of $50\%$ in $\Omegade$ and $\wa$ and $40\%$ in 
$\wzero$ for LSST and is below $40\%$ in all parameters for DES and SNAP 
using 12 bins over the same mass range ($11 < \log(m/\hMsun) < 16$).  
Given our prejudice that the concentration-mass relation should be 
a very smooth function of halo mass, this suggests that self calibration 
can successfully distinguish the variety of such relations realized in 
the universe while providing stringent constraints on dark energy.

The features in Fig.~\ref{fig:selfcal5bin} are qualitatively similar to 
those in Fig.~\ref{fig:selfcalpl}.  The fact that there are multiple 
features in this case reflects the additional freedom in the concentration 
relation.  Each concentration bin influences convergence spectra over 
a particular range of scales.  The largest halos contribute on the largest 
scales, while the smaller halos become increasingly important with increasing 
multipole \cite{CooHuMir00,zhan_knox04}.  In this case, 
the concentration parameters begin to be constrained 
gradually, from highest to lowest mass, as one increases $\ellmax$.  
Aside from these features, the primary aim of Figure~\ref{fig:selfcal5bin} is 
to illustrate the important, global point 
that dark energy parameter constraints are robust to very general 
concentration relations.  Being the most sensitive instrument over the 
range of scales we consider, LSST is most sensitive to these effects, but 
even in this case the cost of avoiding the potential bias shown in 
Fig.~\ref{fig:biases} is low.

\subsection{Constraints on Concentrations from Weak Lensing Tomography}
\label{subsection:constraintc}

To this point, we have considered the unknown structure of halos to be a 
contaminant that degrades our ability to study the cosmic dark energy.  
However, observational constraints on concentrations are interesting in 
their own right.  Indeed, we were driven to consider general 
parameterizations of halo structure 
because the baryonic physics that govern galaxy formation and influence 
the structures of halos is very uncertain.  Direct constraints on 
concentrations may help to inform models of galaxy formation.  
Furthermore, constraints derived from a weak lensing survey 
as described above, would not be subject to the same selection as studies 
of galaxy groups and clusters or studies that 
rely on selecting sample members or stacking systems 
according to some member property.  Therefore, these constraints 
can complement other techniques even if they are not directly 
competitive with other methods.  
We now focus on the constraints on halo concentrations that can be 
derived from weak lensing tomography.

\begin{figure}[t]
\begin{center}
\includegraphics[height=7cm]{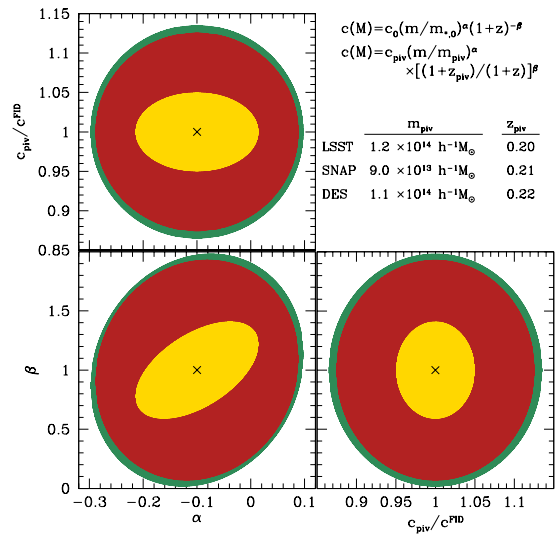}
\caption{
Constraints on the effective halo concentration parameters from tomographic 
weak lensing after marginalizing over the uncertainty in cosmological parameters.  
This plot shows 1$\sigma$ confidence contours on the parameters of the power-law concentration 
relation.  We show the fractional uncertainty in $\cpiv$ relative to the fiducial model.  
Recall that the pivot concentration is the concentration at the halo mass and redshift at 
which concentrations are best determined.  In all panels, from outermost to innermost, 
the contours correspond to those achievable with DES, SNAP, and LSST using  
multipoles up to $\ellmax=3000$.  In the upper right portion of the plot, we list the 
parameterization and the values of the pivot masses and redshifts for convenience.
}
\label{fig:cconpl}
\end{center}
\end{figure}

Consider first constraints on the effective halo mass-concentration relation in 
the power-law concentration mass relation.  Figure~\ref{fig:cconpl} shows 
two-dimensional confidence contours for the concentrations of halos after 
marginalizing over all other parameters, including dark energy.  
For convenience, the pivot masses and redshifts are listed in the 
upper right portion of this plot.  Concentrations are very well 
determined near halo masses of $\sim 1 \times 10^{14} \hMsun$ and a redshift $ z \sim 0.2$.  
The slight differences in these values for each experiment reflect the fact that 
the statistical errors have a different scale dependence for each experiment.  
LSST has the greatest sky coverage, so the LSST pivot mass is the largest.  
Conversely, SNAP has much less sky coverage but with a galaxy number 
density of $\bar{n} = 100$~arcmin$^{-2}$ it is very deep and has the 
smallest $\mpiv$.  These $\mpiv$ and $\zpiv$ values are not unexpected.  It is 
already well known that these masses 
are the largest contributors to the convergence 
power near $\ell \sim 10^{3}$ \cite{CooHuMir00,zhan_knox04}.

The concentration constraints achieved in the course of this 
self calibration are stringent and potentially very useful.  
Marginalizing over the other parameters, 
LSST alone can provide a $\sim 5\%$ constraint on $\cpiv$ at 1$\sigma$, 
and can constrain the mass and redshift power-law indices 
to $\sigma(\alpha)=0.08$ and 
$\sigma(\beta)=0.27$ respectively.  
Both SNAP and DES constrain 
$\cpiv$ at the $\sim 10\%$ level.

\begin{figure}[t]
\begin{center}
\includegraphics[height=7cm]{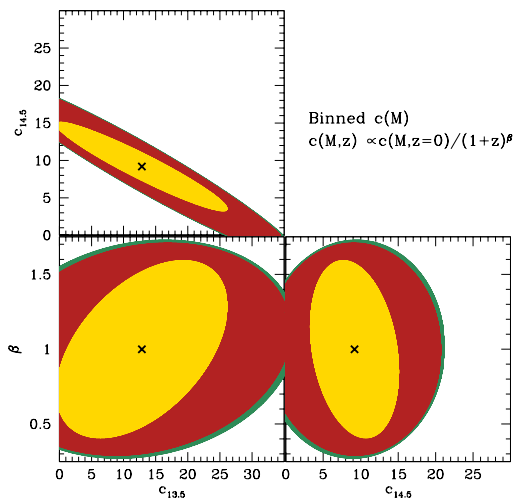}
\caption{
Constraints on the effective halo concentration parameters from tomographic 
weak lensing after marginalizing over the uncertainty in cosmological parameters. 
We show the usefully constrained combination of parameters in the $c_{13.5}$-$c_{14.5}$ 
plane and projections of these parameters with the redshift variable $\beta$.  In 
all panels, from outermost to innermost, the contours correspond to the 
1$\sigma$ contours achievable with DES, SNAP, and LSST with $\ellmax=3000$.
}
\label{fig:c34}
\end{center}
\end{figure}

Figure~\ref{fig:c34} shows concentration parameter constraints on the binned 
$c(m,z)$ relation model.  First, recall that our convention is to designate 
our concentration parameters with a subscript of the value of the logarithm 
of the halo mass at the bin center.  The parameter that describes halo 
concentrations at $m=10^{13.5} \hMsun$ is $c_{13.5}$ and so on.  
Rather than displaying all parameter constraints, Figure~\ref{fig:c34} shows 
a reduced set of parameters that are at least mildly constrained by the experiments 
that we consider.  The other concentration 
parameters are constrained 
at uninteresting levels in all cases.

The fact that halo concentrations are most well constrained near halo masses 
near $\sim 10^{14} \hMsun$ can be seen directly in Figure~\ref{fig:c34}.  This 
is clearly the most well constrained parameter.  In fact, this parameter can 
be constrained with an uncertainty 
$\sigma(c_{14.5})/c_{14.5}=0.48$ relative to the fiducial value 
by LSST alone.  
In the \citet{ma_etal06} fiducial model with greater small-scale power, 
LSST constrains this parameter relatively more stringently with 
$\sigma(c_{14.5})/c_{14.5}=0.31$.  
Generally, this parameter is somewhat degenerate with the 
concentration value at the next lowest mass bin $c_{13.5}$, 
and a combination of these two parameters is constrained at interesting 
levels. As with the power-law case, the redshift dependence is comparably 
poorly constrained.

Though the constraints presented in Fig.~\ref{fig:c34} are modest, 
those in the power-law model of Fig.~\ref{fig:cconpl} are encouraging 
and recall that we have been rather conservative in our priors.  Given 
that precise concentration constraints may inform the modeling of baryonic 
physics, it is interesting to estimate the best possible constraints on 
concentrations that may be achieved with these methods.  To minimize 
the interplay between dark energy and concentration parameters, we 
may assume that dark energy parameters are known extremely well from 
other experiments, for example through the 
supernovae measured by SNAP itself, and that 
we aim to measure effective halo concentrations via weak lensing.  This is 
analogous to the dark energy constraints shown in Fig.~\ref{fig:sigma_std}, 
where it was assumed that concentrations were known perfectly, but in this 
case it is the cosmology that we assume to be well constrained.  

To estimate concentration constraints in this scenario realistically, 
we assume that we have prior knowledge of cosmological parameters that 
exceed the contemporary priors that we have thus far assumed.  To be 
specific, we add priors that are characteristic of projected 
constraints from the Planck cosmic microwave background mission 
and the supernova luminosity distance component of SNAP.  In this 
case, the prior matrix does not have the simple diagonal form of our 
previous prior matrices.  We have taken the prior 
Fisher matrices from the analysis of SNAP and Planck constraints in 
\cite{hu_etal06}.  

\begin{figure}[t]
\begin{center}
\includegraphics[height=7cm]{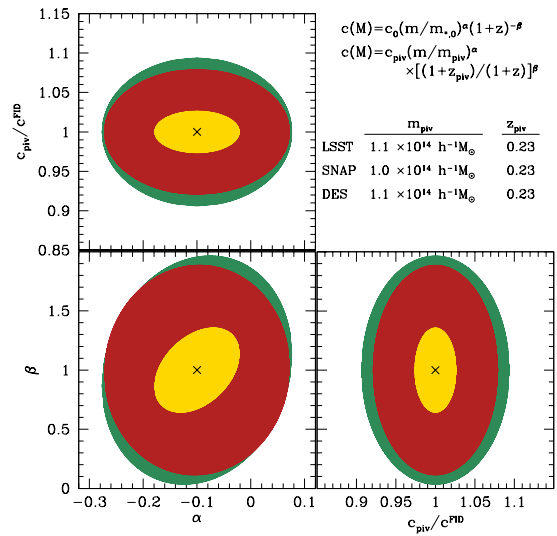}
\caption{
Same as Figure~\ref{fig:cconpl}, but assuming prior constraints on cosmological 
parameters at levels that are expected from a combination of the Planck 
cosmic microwave background satellite and supernovae 
luminosity distance measures from SNAP.  The additional prior matrix for this analysis 
is from Ref.~\cite{hu_etal06}.
}
\label{fig:cconplc}
\end{center}
\end{figure}

We show constraints on effective halo concentrations in the power-law 
concentration model assuming much more restrictive priors on cosmological 
parameters in Figure~\ref{fig:cconplc}.  First, note the slight changes in the 
pivot masses and pivot redshifts.  
More importantly, reducing the dark energy degrees of freedom 
significantly tightens constraints on effective halo concentrations.  For example, 
the 1$\sigma$ LSST constraint on $\cpiv$ drops to about $\sim 2\%$.  
Meanwhile, the LSST constraints on the 
mass dependent power-law index becomes $\sigma(\alpha) \simeq 0.04$.  
Restricting the space of cosmological parameters results in marginal improvement 
on the redshift index, with $\sigma(\beta)\simeq 0.24$.  Both SNAP and DES 
also constrain concentrations stringently near the pivot mass and redshift, 
giving 1$\sigma$ constraints at the $\sim 5\%$ and $\sim 7\%$ levels respectively.

\begin{figure}[t]
\begin{center}
\includegraphics[height=7cm]{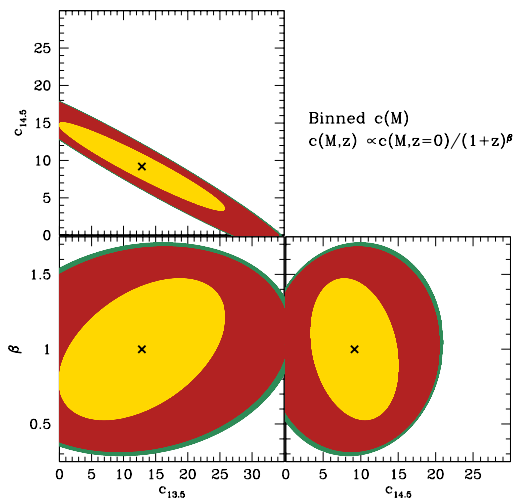}
\caption{
Same as Figure~\ref{fig:c34}, but assuming assuming prior 
constraints on cosmological parameters at levels 
expected from a combination of the Planck cosmic microwave background satellite 
and supernovae luminosity distance measures from SNAP.  The additional 
priors for this analysis are from Ref.~\cite{hu_etal06}.
}
\label{fig:c34c}
\end{center}
\end{figure}

The improvements with better cosmology priors are also 
noteworthy in the case of the general, binned $c(m,z)$ relation.  
We show constraints on the subset of most well constrained parameters in 
Figure~\ref{fig:c34c}.  Consider the case of the LSST experiment, 
for which the constraints are the most stringent.  
In this case, the most significant improvement is 
realized for the redshift index $\beta$, with $\sigma(\beta)=0.3$, while the 
constraint on $c_{13.5}$ decreases to just under $50\%$.  
The constraint on the concentration parameter $c_{14.5}$ from LSST drops to 
rather meaningful levels with $\sigma(c_{14.5})/c_{14.5} \simeq 0.3$.  
Constraints in this regime are 
particularly sensitive to the true cosmology, which sets the signal-to-noise 
level of the forthcoming measurements.  In the \citet{ma_etal06} 
fiducial model with $\sigma_8 \simeq 0.9$, 
the parameters $c_{14.5}$ and $c_{13.5}$ can be independently constrained to 
within $21\%$ and $34\%$ respectively with LSST.

\subsection{Constraint Degradation with Prior Knowledge of Halo Concentrations}
\label{subsection:cpriors}

The experiments we have considered will report results beginning in the year 2011 
in the earliest case of the DES or later.  Of course, it is hard to predict 
the pace of theoretical and observational 
studies of halo concentrations, but it seems reasonable 
to assume that over the next decade there will be some, perhaps significant, 
improvement and that this progress will be able to inform the actual analysis of future 
weak lensing data.  As such, it is interesting to explore the way in which prior constraints on the 
$c(m,z)$ relation influence dark energy parameter degradation from tomographic weak lensing.


\begin{figure}[t]
\begin{center}
\includegraphics[height=9cm]{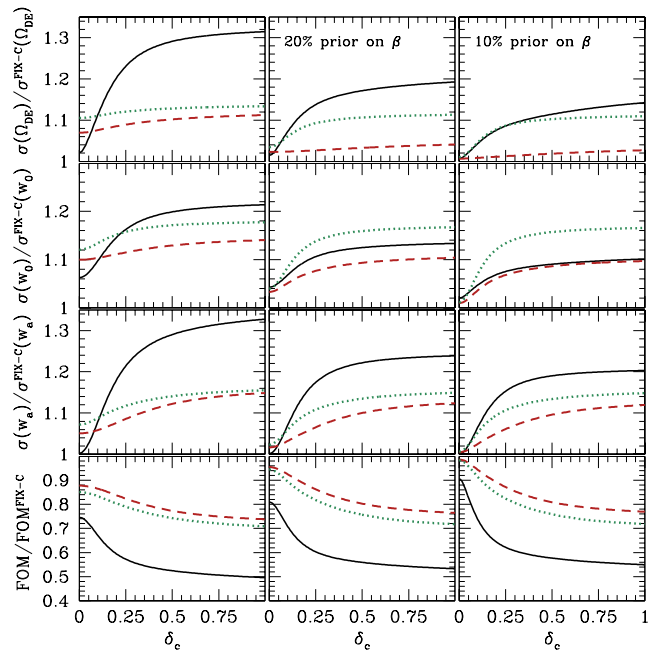}
\caption{
The dependence of dark energy constraints on priors on the $c(m,z)$ relation.  We plot the 
uncertainties in each parameter, scaled by the uncertainty under the assumption of 
perfect knowledge of the concentrations of halos $\sigma^{\mathrm{FIX-C}}$, as a function of 
the fractional prior on the amplitude of concentrations $\delta_{\mathrm{c}}$.  
As in Fig.~\ref{fig:sigma_std}, the rows from top to bottom show constraints on 
$\Omegade$, $\wzero$, and $\wa$, while the bottom panels show the 
dark energy figure of merit, $\fom$.  The leftmost column shows constraints with 
no prior on $\beta$, the middle column shows constraints with a $20\%$ prior on 
$\beta$, and the rightmost column shows constraints with a $10\%$ prior on $\beta$.  
The line types are as in Fig.~\ref{fig:sigma_std}, namely, the {\em solid} line 
gives results for LSST, the {\em dashed} line corresponds to SNAP weak lensing, and 
the {\em dotted} line corresponds to DES.
}
\label{fig:cprior}
\end{center}
\end{figure}

In this section, we consider the influence of priors in 
the case of the binned $c(m,z)$ relation.  For simplicity, we 
assume that the concentrations in all mass bins, $c_{11.5}$ through 
$c_{15.5}$, have a prior constraint of some fraction of their fiducial 
value.  Let us designate this fraction $\delta_{\mathrm{c}}$.  We illustrate 
the effects of prior knowledge of the $c(m,z)$ relation on parameter 
constraints by computing the parameter uncertainties using all information out to 
a maximum multipole $\ellmax=3000$ as a function of prior $\delta_{\mathrm{c}}$.  
Though we take a single parameter to describe the prior in the interest of simplicity, 
it is clearly the prior on the parameter $c_{14.5}$ that is most important 
(see Sec.~\ref{subsection:constraintc}).  
We present results scaled relative to the uncertainties that would be derived 
under the assumption of perfect knowledge of effective halo concentrations, $\sigma^{\mathrm{FIX-C}}$.

The dark energy parameter constraints with prior knowledge of effective concentrations is 
shown in Figure~\ref{fig:cprior}.  As before, LSST is the most sensitive to concentration 
variations and can reap significant benefit from priors on the $c(m,z)$ relation.  In the 
LSST case, prior constraints on the concentration relation at redshift $z=0$ have little 
effect if they are greater than the $\sim 30\%$ level.  This is sensible because LSST 
itself would constrain concentrations in the relevant mass range at this level in the 
absence of prior knowledge.  Naturally, prior constraints on $c(m,z)$ that 
are better than $\sim 30\%$ lead to rapid decreases in parameter 
uncertainties with decreasing $\delta_{\mathrm{c}}$ in each case.

The behavior of parameter constraints as a function of prior reflects the details of 
each of the experiments.  An experiment with enormous sky coverage, such as LSST, can 
establish the amplitude of large-scale density fluctuations 
and the cosmic distance scale with low-multipole 
information and information from the high-redshift tomographic bins 
rather effectively, so that such an experiment can take 
advantage of decreases in $\delta_{\mathrm{c}}$ even with no prior on $\beta$.  This statement 
is less true for SNAP and DES because these experiments have a relatively shorter lever 
arm in multipoles over which to derive constraints.  SNAP is limited by its comparably small 
sky coverage and DES by its comparable shallowness.  As the prior constraints on 
$\beta$ improve, both DES and SNAP become increasingly capable of exploiting prior 
knowledge of the amplitude of the $c(m,z)$ relation as can be seen for 
$\delta_{\mathrm{c}} \lesssim 0.25$ in the two rightmost columns of Fig.~\ref{fig:cprior}.  
In the case of LSST, the figure of merit increases with decreasing $\delta_{\mathrm{c}}$ 
less rapidly than one might suppose based solely on the marginalized uncertainties in 
$\wzero$ and $\wa$ because in this case the correlation between parameters varies strongly 
with $\delta_{\mathrm{c}}$.

In summary, the dark energy constraints from an experiment such as LSST can benefit greatly from 
prior knowledge of the effective concentrations of halos that are better than $\sim 25\%$, 
while constraining the redshift evolution parameter $\beta$, of concentrations to better than 
$\sim 30\%$ 
leads to a rapid decrease in constraints in the absence of prior knowledge of the amplitude 
of the $c(m,z)$ relation.  For DES and SNAP, it is necessary to constrain the redshift evolution 
of $c(m,z)$ before either of these experiments can exploit prior knowledge of the amplitude of 
the $c(m,z)$ relation effectively.  In the cases we study, this requires prior knowledge of the 
$\beta$ parameter to better than $\sim 20\%$.  Knowing the amplitude of the concentration relation 
would reduce parameter degradation to less than $\sim 20\%$ in the case of LSST, while knowing both the 
amplitude and the redshift parameter $\beta$ to less than $\sim 20\%$ would reduce parameter 
degradation to less than $\sim 10\%$ for all experiments.

\section{Summary and Discussion}
\label{section:conc}

We have studied the ability of forthcoming weak lensing surveys to constrain both 
dark energy properties and the concentrations of dark matter halos simultaneously.  
The primary motivations for this work are recent studies demonstrating that 
baryonic physics, neglected in prior forecasting of weak lensing 
constraints on dark energy properties, have a considerable influence on 
predictions for convergence power spectra \cite{jing_etal06,rudd_etal07}.  
Moreover, the large enhancements in convergence spectra predicted by these 
hydrodynamic simulations arose primarily from modifications in the 
internal structures of halos \cite{rudd_etal07}.  Clearly, gravitational 
lensing is sensitive to all matter so here, as in the rest of the paper, 
we refer to halo concentrations, but by this we mean the effective 
concentrations of composite systems comprised of dark matter and baryons 
in radial range $0.04 \lesssim r/\Rvir \lesssim 1$.
We have shown that neglecting these baryonic effects can lead to large 
biases in inferred dark energy parameters unless information from multipoles 
greater than a few hundred is disregarded (see Fig.~\ref{fig:biases}).  
On the other hand, weak lensing surveys have the ability
 to self calibrate for the effects baryons as an alternative 
to eliminating small-scale information.

We modeled the unknown relationship between effective concentration and halo 
mass in two different ways.  In the more restrictive set of models, we considered 
concentration as a power-law function of halo mass.  As we have demonstrated, 
convergence spectra are most sensitive to halo concentrations in a relatively narrow 
range of halo masses near a few $\times 10^{14} \hMsun$, so this class of models may 
not be overly restrictive.  Nevertheless, in currently-favored cosmological 
models, the mass-to-light ratios of halo systems must be a non-monotonic 
function of halo mass, so it is desirable to study relations with more freedom.  Along 
these lines, we studied binned relationships between concentration and halo mass specified 
by average concentration values at a fixed number of halo masses, evenly spaced in 
the logarithm of the halo mass between $11 \le \log(M/\hMsun) \le 16$.  In these models, 
we used spline interpolation to obtain halo concentrations between the values of halo 
mass at which the concentrations were specified.  Throughout this paper, we have presented 
results for a concentration-mass relationship specified at five bins in this mass interval, 
but have verified that our results are relatively insensitive to the number of bins up to 
about twelve bins.  Coupled with our prejudice that these effective concentrations will 
likely be very smooth functions of halo mass, this class of model is likely to have 
sufficient flexibility to produce the types of 
mass-concentration relations that are realized in nature.  
In both cases, we allowed for a redshift dependence to the modification 
in the halo mass-concentration relation by 
taking $c(m,z)=c(m,z=0)/(1+z)^{\beta}$.

Our study implies that self-calibration is a quite promising 
alternative to  simply eliminating multipoles beyond which significant 
parameter biases may be likely.  
Comparing constraints obtained by removing multipoles beyond a few 
hundred with constraints that include all multipoles up to $\ell=3000$ and assume perfect 
knowledge of halo concentrations, the degradation in parameter uncertainties 
is a factor of $\sim 1.5-3$.  The details of the degradation 
depend upon the parameter of interest, the experiment, and 
the fiducial cosmological model.  As it is the most sensitive of the forthcoming instruments 
on large scales, the requirements for LSST to be unbiased are the most restrictive and 
indicate that multipoles beyond $\sim 300$ should not be considered in parameter extraction.  
The corresponding degradation in parameter constraints for LSST is a factor of $\sim 3$ for 
all parameters.  On the contrary, our study suggests that self calibration is a more 
fruitful alternative.  Even in our most permissive concentration models, dark energy 
parameter constraints from this self-calibration exercise are never degraded by more than 
$36\%$ and more restrictive models (if they can be justified) do significantly better.  
In particular, if the concentration-mass relation can be constrained 
to better than $20\%$ at a specific redshift, 
the constraints derived from an LSST-like experiment 
will be degraded by only $\sim 20\%$.  If, in addition, the power-law index describing 
the redshift evolution of effective halo concentrations is also known to within 
$20\%$, then dark energy parameter constraints will be degraded by no more than 
$\sim 10\%$ for any of LSST, DES, or SNAP individually.

Self calibration as an alternative to removing the information contained in high-$\ell$ 
multipoles has another advantage; it provides direct constraints on 
halo concentrations that may complement, if not be competitive with, other methods.  
In the case of the power-law class of models for effective halo concentrations as a 
function of mass and redshift, 
LSST alone can constrain halo concentrations near the pivot mass, $\mpiv \sim 10^{14} \hMsun$, 
and redshift, $\zpiv \sim 0.2$, at the $\sim 5\%$ level with only modest 
prior constraints on cosmology that are typical of our contemporary 
knowledge of cosmological parameters.  Both DES and SNAP constrain 
halo concentrations at the pivot mass and redshift to about $\sim 10\%$.  
Generalizing to the case of the arbitrary, binned relation for halo concentrations, 
LSST alone constrains halo concentrations near $m \sim 3 \times 10^{14} \hMsun$ 
to within $48\%$.  Adding prior constraints on cosmological parameters from 
the Planck cosmic microwave background satellite and supernova luminosity distances 
as measured from SNAP reduce this to a $30\%$ constraint.  Though modest, this 
constraint would not be subject to the same systematics and selection issues 
as other techniques.

Our work has several shortcomings.  Of course, our framework for all of these calculations 
has been the halo model, and this phenomenological model has known inadequacies.  More 
importantly, we have modeled the effect of baryons as entirely restricted to the internal 
structures of halos.  Though this is a good approximation in current simulations \cite{rudd_etal07}, 
it is unreasonable to expect that this is strictly true and no study has yet 
addressed potentially-important, though sub-dominant, additional modification to 
matter power spectra.  Changes to the detailed profile shapes, and the 
mass function of halos (already measured in Ref.~\cite{rudd_etal07}), 
the clustering of halos are all likely to be important at some level.  
To address this issue is an ambitious goal which lies 
well beyond the scope of our study, yet it is a goal that must be achieved in order 
to take full advantage of these forthcoming experiments.  The degree to which 
baryons will modify the matter power spectra will likely remain uncertain for 
some time, yet we must at least enumerate the different ways in which 
the power spectra are modified.  As an example, the study of Ref.~\cite{rudd_etal07} 
does show percent-level residuals between their simulation results and the results of 
a model based entirely on modified halo concentrations.  Although the results 
of Ref.~\cite{rudd_etal07} are noisy due to their small simulation volume, 
we can use their data to obtain a first guess at the residual 
bias that may remain after allowing for variation in halo concentration parameters.  
Our estimates suggest that in the case of LSST the bias is a few percent of 
the $1\sigma$ parameter uncertainty at $\ellmax=10^{3}$ and is of order 
$| \delta |/\sigma \simeq 0.1$ at $\ellmax=3000$ for both $\wzero$ and $\wa$.  
The residual biases in the case of DES and SNAP are smaller yet.  
Furthermore, we have described forthcoming experiments in a simple manner and 
so detailed comparisons between the projected performances of instruments 
based on this study should not be regarded as definitive.  
Certainly, systematic issues regarding shape measurements will be 
better controlled in space-borne instruments like SNAP than from the 
ground.  Moreover, SNAP with potentially much greater depth than either 
DES or LSST will be sensitive to relatively higher-redshift, larger-scale 
fluctuations at a fixed angular scale.  On such scales, the fluctuation 
spectrum will be less subject to nonlinear effects.  
Lastly, we have ignored all additional systematics whether theoretical 
or observational.  We anticipate that 
in the years leading up to the surveys we have studied, a considerable 
amount of effort will be put into making theoretical predictions 
more accurate.  We expect --- or at least hope --- that more 
sophisticated methods for predicting matter power spectra will 
supplant current techniques and find it most likely that the 
parameterizations we adopted will not be those that are implemented 
in the actual data analyses.  Nevertheless, our study gives a preliminary 
indication that calibration of uncertain baryonic effects will be 
possible at minimal cost and that this calibration may provide 
interesting information about halo structure that may be used 
to diagnose models of the evolution of the baryons themselves.

In the context of our models, calibrating effective halo concentrations 
in parallel with deriving constraints on dark energy parameters 
is extremely successful and the preceding paragraph is 
not intended to undercut our primary results and conclusions.  
We are optimistic that this success will carry over to more complex models that may 
include systematics and additional modifications to power spectra 
due to baryonic processes.  Part of our optimism stems directly from 
the success with which halo structure can be calibrated, as we have 
demonstrated here.  In addition, we are optimistic because we have used 
only a fraction of the information contained within these surveys.  In 
particular, we restricted our analysis to multipoles $\ell < 3000$ in 
order to remain in the regime where Gaussian statistics and other weak 
lensing approximations are valid.  Accounting for non-Gaussianity and 
dropping additional assumptions, such as the Born approximation and the 
use of reduced shear \cite{dodelson_etal06}, would 
allow higher multipoles to be included in the analysis.  This would 
greatly increase the constraining power of these week lensing surveys.  More specifically, 
SNAP is the deepest survey we have considered by a large margin.  As such, SNAP 
suffers the most from setting $\ellmax=3000$ and going beyond this limit would enable 
SNAP to outperform the projections we present here (while systematics will also be 
better controlled from space).  An important implication of this is that 
upon extending the analysis to small scales, 
SNAP would provide the best determination of baryonic effects, eliminating degeneracies 
that may limit dark energy constraints.  Of course, gains from small-scale observations 
would also come with the challenge of predicting power spectra out 
to higher multipoles.  Furthermore, 
we have used only weak lensing shear information and have not 
utilized the galaxy power spectra and galaxy-shear cross spectra 
that would also be contained within these surveys and would make 
these constraints more robust \cite{Bernstein:2003es,hu_jain04}.  
Our study shows the 
great promise of future surveys to constrain dark energy along with 
numerous other phenomena, but is is only the tip of the iceberg and 
we must meet several theoretical challenges in order to exploit the 
enormous amount of information that will be available in forthcoming 
photometric surveys.

\begin{acknowledgments}
We are grateful to Gary Bernstein, Dragan Huterer, Andrey Kravtsov, 
Zhaoming Ma, Brant Robertson, Chaz Shapiro, Louie Strigari, Jeremy Tinker, 
Sheng Wang, Martin White, and Hu Zhan for many useful discussions.  
This work was supported by the Kavli Institute for Cosmological 
Physics at The University of Chicago under the NSF Physics Frontier Grant 
NSF PHY 0114422.
ARZ has been additionally funded by the University of Pittsburgh and
the National Science Foundation (NSF) Astronomy and Astrophysics 
Postdoctoral Fellowship program through grant AST 0602122.
  WH was additionally supported by the
U.S.~Dept. of Energy contract DE-FG02-90ER-40560 and
 the David and Lucile Packard Foundation. 
This work made use of the 
National Aeronautics and Space Administration Astrophysics Data 
System.
\end{acknowledgments}

\bibliography{wlc}

\begin{thebibliography}{84}
\expandafter\ifx\csname natexlab\endcsname\relax\def\natexlab#1{#1}\fi
\expandafter\ifx\csname bibnamefont\endcsname\relax
  \def\bibnamefont#1{#1}\fi
\expandafter\ifx\csname bibfnamefont\endcsname\relax
  \def\bibfnamefont#1{#1}\fi
\expandafter\ifx\csname citenamefont\endcsname\relax
  \def\citenamefont#1{#1}\fi
\expandafter\ifx\csname url\endcsname\relax
  \def\url#1{\texttt{#1}}\fi
\expandafter\ifx\csname urlprefix\endcsname\relax\def\urlprefix{URL }\fi
\providecommand{\bibinfo}[2]{#2}
\providecommand{\eprint}[2][]{\url{#2}}

\bibitem[{\citenamefont{{Riess} et~al.}(1998)\citenamefont{{Riess},
  {Filippenko}, {Challis}, and {et al.}}}]{riess_etal98}
\bibinfo{author}{\bibfnamefont{A.~G.} \bibnamefont{{Riess}}},
  \bibinfo{author}{\bibfnamefont{A.~V.} \bibnamefont{{Filippenko}}},
  \bibinfo{author}{\bibfnamefont{P.}~\bibnamefont{{Challis}}},
  \bibnamefont{and} \bibinfo{author}{\bibnamefont{{et al.}}},
  \bibinfo{journal}{\aj} \textbf{\bibinfo{volume}{116}}, \bibinfo{pages}{1009}
  (\bibinfo{year}{1998}), \eprint{astro-ph/9805201}.

\bibitem[{\citenamefont{{Perlmutter} et~al.}(1999)\citenamefont{{Perlmutter},
  {Aldering}, {Goldhaber}, and {et al.}}}]{perlmutter_etal99}
\bibinfo{author}{\bibfnamefont{S.}~\bibnamefont{{Perlmutter}}},
  \bibinfo{author}{\bibfnamefont{G.}~\bibnamefont{{Aldering}}},
  \bibinfo{author}{\bibfnamefont{G.}~\bibnamefont{{Goldhaber}}},
  \bibnamefont{and} \bibinfo{author}{\bibnamefont{{et al.}}},
  \bibinfo{journal}{\apj} \textbf{\bibinfo{volume}{517}}, \bibinfo{pages}{565}
  (\bibinfo{year}{1999}), \eprint{astro-ph/9812133}.

\bibitem[{\citenamefont{{Tegmark} et~al.}(2004)\citenamefont{{Tegmark},
  {Strauss}, {Blanton}, and {et al.}}}]{tegmark_etal04}
\bibinfo{author}{\bibfnamefont{M.}~\bibnamefont{{Tegmark}}},
  \bibinfo{author}{\bibfnamefont{M.~A.} \bibnamefont{{Strauss}}},
  \bibinfo{author}{\bibfnamefont{M.~R.} \bibnamefont{{Blanton}}},
  \bibnamefont{and} \bibinfo{author}{\bibnamefont{{et al.}}},
  \bibinfo{journal}{\prd} \textbf{\bibinfo{volume}{69}},
  \bibinfo{pages}{103501} (\bibinfo{year}{2004}), \eprint{astro-ph/0310723}.

\bibitem[{\citenamefont{{Riess} et~al.}(2004)\citenamefont{{Riess}, {Strolger},
  {Tonry}, {Casertano}, {Ferguson}, {Mobasher}, {Challis}, {Filippenko}, {Jha},
  {Li} et~al.}}]{riess_etal04}
\bibinfo{author}{\bibfnamefont{A.~G.} \bibnamefont{{Riess}}},
  \bibinfo{author}{\bibfnamefont{L.-G.} \bibnamefont{{Strolger}}},
  \bibinfo{author}{\bibfnamefont{J.}~\bibnamefont{{Tonry}}},
  \bibinfo{author}{\bibfnamefont{S.}~\bibnamefont{{Casertano}}},
  \bibinfo{author}{\bibfnamefont{H.~C.} \bibnamefont{{Ferguson}}},
  \bibinfo{author}{\bibfnamefont{B.}~\bibnamefont{{Mobasher}}},
  \bibinfo{author}{\bibfnamefont{P.}~\bibnamefont{{Challis}}},
  \bibinfo{author}{\bibfnamefont{A.~V.} \bibnamefont{{Filippenko}}},
  \bibinfo{author}{\bibfnamefont{S.}~\bibnamefont{{Jha}}},
  \bibinfo{author}{\bibfnamefont{W.}~\bibnamefont{{Li}}}, \bibnamefont{et~al.},
  \bibinfo{journal}{\apj} \textbf{\bibinfo{volume}{607}}, \bibinfo{pages}{665}
  (\bibinfo{year}{2004}), \eprint{astro-ph/0402512}.

\bibitem[{\citenamefont{{Eisenstein} et~al.}(2005)\citenamefont{{Eisenstein},
  {Zehavi}, {Hogg}, and {et al.}}}]{eisenstein_etal05}
\bibinfo{author}{\bibfnamefont{D.~J.} \bibnamefont{{Eisenstein}}},
  \bibinfo{author}{\bibfnamefont{I.}~\bibnamefont{{Zehavi}}},
  \bibinfo{author}{\bibfnamefont{D.~W.} \bibnamefont{{Hogg}}},
  \bibnamefont{and} \bibinfo{author}{\bibnamefont{{et al.}}},
  \bibinfo{journal}{\apj} \textbf{\bibinfo{volume}{633}}, \bibinfo{pages}{560}
  (\bibinfo{year}{2005}), \eprint{astro-ph/0501171}.

\bibitem[{\citenamefont{{Spergel} et~al.}(2007)\citenamefont{{Spergel}, {Bean},
  {Dor{\'e}}, {Nolta}, {Bennett}, {Dunkley}, {Hinshaw}, {Jarosik}, {Komatsu},
  {Page} et~al.}}]{spergel_etal07}
\bibinfo{author}{\bibfnamefont{D.~N.} \bibnamefont{{Spergel}}},
  \bibinfo{author}{\bibfnamefont{R.}~\bibnamefont{{Bean}}},
  \bibinfo{author}{\bibfnamefont{O.}~\bibnamefont{{Dor{\'e}}}},
  \bibinfo{author}{\bibfnamefont{M.~R.} \bibnamefont{{Nolta}}},
  \bibinfo{author}{\bibfnamefont{C.~L.} \bibnamefont{{Bennett}}},
  \bibinfo{author}{\bibfnamefont{J.}~\bibnamefont{{Dunkley}}},
  \bibinfo{author}{\bibfnamefont{G.}~\bibnamefont{{Hinshaw}}},
  \bibinfo{author}{\bibfnamefont{N.}~\bibnamefont{{Jarosik}}},
  \bibinfo{author}{\bibfnamefont{E.}~\bibnamefont{{Komatsu}}},
  \bibinfo{author}{\bibfnamefont{L.}~\bibnamefont{{Page}}},
  \bibnamefont{et~al.}, \bibinfo{journal}{\apjs}
  \textbf{\bibinfo{volume}{170}}, \bibinfo{pages}{377} (\bibinfo{year}{2007}),
  \eprint{arXiv:astro-ph/0603449}.

\bibitem[{\citenamefont{{Tegmark} et~al.}(2006)\citenamefont{{Tegmark},
  {Eisenstein}, {Strauss}, and {et al.}}}]{tegmark_etal06}
\bibinfo{author}{\bibfnamefont{M.}~\bibnamefont{{Tegmark}}},
  \bibinfo{author}{\bibfnamefont{D.~J.} \bibnamefont{{Eisenstein}}},
  \bibinfo{author}{\bibfnamefont{M.~A.} \bibnamefont{{Strauss}}},
  \bibnamefont{and} \bibinfo{author}{\bibnamefont{{et al.}}},
  \bibinfo{journal}{\prd} \textbf{\bibinfo{volume}{74}},
  \bibinfo{pages}{123507} (\bibinfo{year}{2006}), \eprint{astro-ph/0608632}.

\bibitem[{\citenamefont{{Astier} et~al.}(2006)\citenamefont{{Astier}, {Guy},
  {Regnault}, and {et al.}}}]{astier_etal06}
\bibinfo{author}{\bibfnamefont{P.}~\bibnamefont{{Astier}}},
  \bibinfo{author}{\bibfnamefont{J.}~\bibnamefont{{Guy}}},
  \bibinfo{author}{\bibfnamefont{N.}~\bibnamefont{{Regnault}}},
  \bibnamefont{and} \bibinfo{author}{\bibnamefont{{et al.}}},
  \bibinfo{journal}{\aap} \textbf{\bibinfo{volume}{447}}, \bibinfo{pages}{31}
  (\bibinfo{year}{2006}), \eprint{astro-ph/0510447}.

\bibitem[{\citenamefont{{Wood-Vasey} et~al.}(2007)\citenamefont{{Wood-Vasey},
  {Miknaitis}, {Stubbs}, and {et al.}}}]{wood-vasey_etal07}
\bibinfo{author}{\bibfnamefont{W.~M.} \bibnamefont{{Wood-Vasey}}},
  \bibinfo{author}{\bibfnamefont{G.}~\bibnamefont{{Miknaitis}}},
  \bibinfo{author}{\bibfnamefont{C.~W.} \bibnamefont{{Stubbs}}},
  \bibnamefont{and} \bibinfo{author}{\bibnamefont{{et al.}}},
  \bibinfo{journal}{{\apj~submitted} (astro-ph/0701041)}
  (\bibinfo{year}{2007}), \eprint{astro-ph/0701041}.

\bibitem[{\citenamefont{{Hu} and {Tegmark}}(1999)}]{hu_tegmark99}
\bibinfo{author}{\bibfnamefont{W.}~\bibnamefont{{Hu}}} \bibnamefont{and}
  \bibinfo{author}{\bibfnamefont{M.}~\bibnamefont{{Tegmark}}},
  \bibinfo{journal}{\apjl} \textbf{\bibinfo{volume}{514}}, \bibinfo{pages}{L65}
  (\bibinfo{year}{1999}), \eprint{astro-ph/9811168}.

\bibitem[{\citenamefont{{Hu}}(1999)}]{hu99}
\bibinfo{author}{\bibfnamefont{W.}~\bibnamefont{{Hu}}},
  \bibinfo{journal}{\apjl} \textbf{\bibinfo{volume}{522}}, \bibinfo{pages}{L21}
  (\bibinfo{year}{1999}), \eprint{astro-ph/9904153}.

\bibitem[{\citenamefont{{Huterer}}(2002)}]{huterer02}
\bibinfo{author}{\bibfnamefont{D.}~\bibnamefont{{Huterer}}},
  \bibinfo{journal}{\prd} \textbf{\bibinfo{volume}{65}},
  \bibinfo{pages}{063001} (\bibinfo{year}{2002}), \eprint{astro-ph/0106399}.

\bibitem[{\citenamefont{{Heavens}}(2003)}]{heavens03}
\bibinfo{author}{\bibfnamefont{A.}~\bibnamefont{{Heavens}}},
  \bibinfo{journal}{\mnras} \textbf{\bibinfo{volume}{343}},
  \bibinfo{pages}{1327} (\bibinfo{year}{2003}), \eprint{astro-ph/0304151}.

\bibitem[{\citenamefont{{Refregier}}(2003)}]{refregier03}
\bibinfo{author}{\bibfnamefont{A.}~\bibnamefont{{Refregier}}},
  \bibinfo{journal}{\araa} \textbf{\bibinfo{volume}{41}}, \bibinfo{pages}{645}
  (\bibinfo{year}{2003}), \eprint{astro-ph/0307212}.

\bibitem[{\citenamefont{{Refregier} et~al.}(2004)\citenamefont{{Refregier},
  {Massey}, {Rhodes}, {Ellis}, {Albert}, {Bacon}, {Bernstein}, {McKay}, and
  {Perlmutter}}}]{refregier_etal04}
\bibinfo{author}{\bibfnamefont{A.}~\bibnamefont{{Refregier}}},
  \bibinfo{author}{\bibfnamefont{R.}~\bibnamefont{{Massey}}},
  \bibinfo{author}{\bibfnamefont{J.}~\bibnamefont{{Rhodes}}},
  \bibinfo{author}{\bibfnamefont{R.}~\bibnamefont{{Ellis}}},
  \bibinfo{author}{\bibfnamefont{J.}~\bibnamefont{{Albert}}},
  \bibinfo{author}{\bibfnamefont{D.}~\bibnamefont{{Bacon}}},
  \bibinfo{author}{\bibfnamefont{G.}~\bibnamefont{{Bernstein}}},
  \bibinfo{author}{\bibfnamefont{T.}~\bibnamefont{{McKay}}}, \bibnamefont{and}
  \bibinfo{author}{\bibfnamefont{S.}~\bibnamefont{{Perlmutter}}},
  \bibinfo{journal}{\aj} \textbf{\bibinfo{volume}{127}}, \bibinfo{pages}{3102}
  (\bibinfo{year}{2004}), \eprint{astro-ph/0304419}.

\bibitem[{\citenamefont{{Song} and {Knox}}(2004)}]{song_knox04}
\bibinfo{author}{\bibfnamefont{Y.-S.} \bibnamefont{{Song}}} \bibnamefont{and}
  \bibinfo{author}{\bibfnamefont{L.}~\bibnamefont{{Knox}}},
  \bibinfo{journal}{\prd} \textbf{\bibinfo{volume}{70}},
  \bibinfo{pages}{063510} (\bibinfo{year}{2004}), \eprint{astro-ph/0312175}.

\bibitem[{\citenamefont{{Takada} and {Jain}}(2004)}]{takada_jain04}
\bibinfo{author}{\bibfnamefont{M.}~\bibnamefont{{Takada}}} \bibnamefont{and}
  \bibinfo{author}{\bibfnamefont{B.}~\bibnamefont{{Jain}}},
  \bibinfo{journal}{\mnras} \textbf{\bibinfo{volume}{348}},
  \bibinfo{pages}{897} (\bibinfo{year}{2004}), \eprint{astro-ph/0310125}.

\bibitem[{\citenamefont{{Takada} and {White}}(2004)}]{takada_white04}
\bibinfo{author}{\bibfnamefont{M.}~\bibnamefont{{Takada}}} \bibnamefont{and}
  \bibinfo{author}{\bibfnamefont{M.}~\bibnamefont{{White}}},
  \bibinfo{journal}{\apjl} \textbf{\bibinfo{volume}{601}}, \bibinfo{pages}{L1}
  (\bibinfo{year}{2004}), \eprint{astro-ph/0311104}.

\bibitem[{\citenamefont{{Dodelson} and {Zhang}}(2005)}]{dodelson_zhang05}
\bibinfo{author}{\bibfnamefont{S.}~\bibnamefont{{Dodelson}}} \bibnamefont{and}
  \bibinfo{author}{\bibfnamefont{P.}~\bibnamefont{{Zhang}}},
  \bibinfo{journal}{\prd} \textbf{\bibinfo{volume}{72}},
  \bibinfo{pages}{083001} (\bibinfo{year}{2005}), \eprint{astro-ph/0501063}.

\bibitem[{\citenamefont{{Albrecht}
  et~al.}(2006{\natexlab{a}})\citenamefont{{Albrecht}, {Bernstein}, {Cahn},
  {Freedman}, {Hewitt}, {Hu}, {Huth}, {Kamionkowski}, {Kolb}, {Knox}
  et~al.}}]{albrecht_etal06}
\bibinfo{author}{\bibfnamefont{A.}~\bibnamefont{{Albrecht}}},
  \bibinfo{author}{\bibfnamefont{G.}~\bibnamefont{{Bernstein}}},
  \bibinfo{author}{\bibfnamefont{R.}~\bibnamefont{{Cahn}}},
  \bibinfo{author}{\bibfnamefont{W.~L.} \bibnamefont{{Freedman}}},
  \bibinfo{author}{\bibfnamefont{J.}~\bibnamefont{{Hewitt}}},
  \bibinfo{author}{\bibfnamefont{W.}~\bibnamefont{{Hu}}},
  \bibinfo{author}{\bibfnamefont{J.}~\bibnamefont{{Huth}}},
  \bibinfo{author}{\bibfnamefont{M.}~\bibnamefont{{Kamionkowski}}},
  \bibinfo{author}{\bibfnamefont{E.~W.} \bibnamefont{{Kolb}}},
  \bibinfo{author}{\bibfnamefont{L.}~\bibnamefont{{Knox}}},
  \bibnamefont{et~al.}, \bibinfo{journal}{(astro-ph/0609591)}
  (\bibinfo{year}{2006}{\natexlab{a}}), \eprint{astro-ph/0609591}.

\bibitem[{\citenamefont{{Zhan}}(2006)}]{zhan06}
\bibinfo{author}{\bibfnamefont{H.}~\bibnamefont{{Zhan}}},
  \bibinfo{journal}{Journal of Cosmology and Astro-Particle Physics}
  \textbf{\bibinfo{volume}{8}}, \bibinfo{pages}{8} (\bibinfo{year}{2006}),
  \eprint{astro-ph/0605696}.

\bibitem[{\citenamefont{{Huterer} and {Takada}}(2005)}]{huterer_takada05}
\bibinfo{author}{\bibfnamefont{D.}~\bibnamefont{{Huterer}}} \bibnamefont{and}
  \bibinfo{author}{\bibfnamefont{M.}~\bibnamefont{{Takada}}},
  \bibinfo{journal}{Astroparticle Physics} \textbf{\bibinfo{volume}{23}},
  \bibinfo{pages}{369} (\bibinfo{year}{2005}), \eprint{astro-ph/0412142}.

\bibitem[{\citenamefont{{Huterer} et~al.}(2006)\citenamefont{{Huterer},
  {Takada}, {Bernstein}, and {Jain}}}]{huterer_etal06}
\bibinfo{author}{\bibfnamefont{D.}~\bibnamefont{{Huterer}}},
  \bibinfo{author}{\bibfnamefont{M.}~\bibnamefont{{Takada}}},
  \bibinfo{author}{\bibfnamefont{G.}~\bibnamefont{{Bernstein}}},
  \bibnamefont{and} \bibinfo{author}{\bibfnamefont{B.}~\bibnamefont{{Jain}}},
  \bibinfo{journal}{\mnras} \textbf{\bibinfo{volume}{366}},
  \bibinfo{pages}{101} (\bibinfo{year}{2006}), \eprint{astro-ph/0506030}.

\bibitem[{\citenamefont{{Heitmann} et~al.}(2005)\citenamefont{{Heitmann},
  {Ricker}, {Warren}, and {Habib}}}]{heitmann_etal05}
\bibinfo{author}{\bibfnamefont{K.}~\bibnamefont{{Heitmann}}},
  \bibinfo{author}{\bibfnamefont{P.~M.} \bibnamefont{{Ricker}}},
  \bibinfo{author}{\bibfnamefont{M.~S.} \bibnamefont{{Warren}}},
  \bibnamefont{and} \bibinfo{author}{\bibfnamefont{S.}~\bibnamefont{{Habib}}},
  \bibinfo{journal}{\apjs} \textbf{\bibinfo{volume}{160}}, \bibinfo{pages}{28}
  (\bibinfo{year}{2005}), \eprint{astro-ph/0411795}.

\bibitem[{\citenamefont{{White}}(2004)}]{white04}
\bibinfo{author}{\bibfnamefont{M.}~\bibnamefont{{White}}},
  \bibinfo{journal}{Astroparticle Physics} \textbf{\bibinfo{volume}{22}},
  \bibinfo{pages}{211} (\bibinfo{year}{2004}), \eprint{astro-ph/0405593}.

\bibitem[{\citenamefont{{Zhan} and {Knox}}(2004)}]{zhan_knox04}
\bibinfo{author}{\bibfnamefont{H.}~\bibnamefont{{Zhan}}} \bibnamefont{and}
  \bibinfo{author}{\bibfnamefont{L.}~\bibnamefont{{Knox}}},
  \bibinfo{journal}{\apjl} \textbf{\bibinfo{volume}{616}}, \bibinfo{pages}{L75}
  (\bibinfo{year}{2004}), \eprint{astro-ph/0409198}.

\bibitem[{\citenamefont{{Jing} et~al.}(2006)\citenamefont{{Jing}, {Zhang},
  {Lin}, {Gao}, and {Springel}}}]{jing_etal06}
\bibinfo{author}{\bibfnamefont{Y.~P.} \bibnamefont{{Jing}}},
  \bibinfo{author}{\bibfnamefont{P.}~\bibnamefont{{Zhang}}},
  \bibinfo{author}{\bibfnamefont{W.~P.} \bibnamefont{{Lin}}},
  \bibinfo{author}{\bibfnamefont{L.}~\bibnamefont{{Gao}}}, \bibnamefont{and}
  \bibinfo{author}{\bibfnamefont{V.}~\bibnamefont{{Springel}}},
  \bibinfo{journal}{\apjl} \textbf{\bibinfo{volume}{640}},
  \bibinfo{pages}{L119} (\bibinfo{year}{2006}), \eprint{astro-ph/0512426}.

\bibitem[{\citenamefont{{Rudd} et~al.}(2007)\citenamefont{{Rudd}, {Zentner},
  and {Kravtsov}}}]{rudd_etal07}
\bibinfo{author}{\bibfnamefont{D.~H.} \bibnamefont{{Rudd}}},
  \bibinfo{author}{\bibfnamefont{A.~R.} \bibnamefont{{Zentner}}},
  \bibnamefont{and} \bibinfo{author}{\bibfnamefont{A.~V.}
  \bibnamefont{{Kravtsov}}}, \bibinfo{journal}{\apj~Submitted}
  (\bibinfo{year}{2007}), \eprint{astro-ph/0703741}.

\bibitem[{\citenamefont{{Sheldon} et~al.}(2004)\citenamefont{{Sheldon},
  {Johnston}, {Frieman}, {Scranton}, {McKay}, {Connolly}, {Budav{\'a}ri},
  {Zehavi}, {Bahcall}, {Brinkmann} et~al.}}]{sheldon_etal04}
\bibinfo{author}{\bibfnamefont{E.~S.} \bibnamefont{{Sheldon}}},
  \bibinfo{author}{\bibfnamefont{D.~E.} \bibnamefont{{Johnston}}},
  \bibinfo{author}{\bibfnamefont{J.~A.} \bibnamefont{{Frieman}}},
  \bibinfo{author}{\bibfnamefont{R.}~\bibnamefont{{Scranton}}},
  \bibinfo{author}{\bibfnamefont{T.~A.} \bibnamefont{{McKay}}},
  \bibinfo{author}{\bibfnamefont{A.~J.} \bibnamefont{{Connolly}}},
  \bibinfo{author}{\bibfnamefont{T.}~\bibnamefont{{Budav{\'a}ri}}},
  \bibinfo{author}{\bibfnamefont{I.}~\bibnamefont{{Zehavi}}},
  \bibinfo{author}{\bibfnamefont{N.~A.} \bibnamefont{{Bahcall}}},
  \bibinfo{author}{\bibfnamefont{J.}~\bibnamefont{{Brinkmann}}},
  \bibnamefont{et~al.}, \bibinfo{journal}{\aj} \textbf{\bibinfo{volume}{127}},
  \bibinfo{pages}{2544} (\bibinfo{year}{2004}),
  \eprint{arXiv:astro-ph/0312036}.

\bibitem[{\citenamefont{{{\L}okas} et~al.}(2006)\citenamefont{{{\L}okas},
  {Wojtak}, {Gottl{\"o}ber}, {Mamon}, and {Prada}}}]{lokas_etal06}
\bibinfo{author}{\bibfnamefont{E.~L.} \bibnamefont{{{\L}okas}}},
  \bibinfo{author}{\bibfnamefont{R.}~\bibnamefont{{Wojtak}}},
  \bibinfo{author}{\bibfnamefont{S.}~\bibnamefont{{Gottl{\"o}ber}}},
  \bibinfo{author}{\bibfnamefont{G.~A.} \bibnamefont{{Mamon}}},
  \bibnamefont{and} \bibinfo{author}{\bibfnamefont{F.}~\bibnamefont{{Prada}}},
  \bibinfo{journal}{\mnras} \textbf{\bibinfo{volume}{367}},
  \bibinfo{pages}{1463} (\bibinfo{year}{2006}),
  \eprint{arXiv:astro-ph/0511723}.

\bibitem[{\citenamefont{{Mandelbaum}
  et~al.}(2006{\natexlab{a}})\citenamefont{{Mandelbaum}, {Seljak}, {Cool},
  {Blanton}, {Hirata}, and {Brinkmann}}}]{mandelbaum_etal06}
\bibinfo{author}{\bibfnamefont{R.}~\bibnamefont{{Mandelbaum}}},
  \bibinfo{author}{\bibfnamefont{U.}~\bibnamefont{{Seljak}}},
  \bibinfo{author}{\bibfnamefont{R.~J.} \bibnamefont{{Cool}}},
  \bibinfo{author}{\bibfnamefont{M.}~\bibnamefont{{Blanton}}},
  \bibinfo{author}{\bibfnamefont{C.~M.} \bibnamefont{{Hirata}}},
  \bibnamefont{and}
  \bibinfo{author}{\bibfnamefont{J.}~\bibnamefont{{Brinkmann}}},
  \bibinfo{journal}{\mnras} \textbf{\bibinfo{volume}{372}},
  \bibinfo{pages}{758} (\bibinfo{year}{2006}{\natexlab{a}}),
  \eprint{arXiv:astro-ph/0605476}.

\bibitem[{\citenamefont{{Buote} et~al.}(2006)\citenamefont{{Buote},
  {Gastaldello}, {Humphrey}, {Zappacosta}, {Bullock}, {Brighenti}, and
  {Mathews}}}]{buote_etal06}
\bibinfo{author}{\bibfnamefont{D.~A.} \bibnamefont{{Buote}}},
  \bibinfo{author}{\bibfnamefont{F.}~\bibnamefont{{Gastaldello}}},
  \bibinfo{author}{\bibfnamefont{P.~J.} \bibnamefont{{Humphrey}}},
  \bibinfo{author}{\bibfnamefont{L.}~\bibnamefont{{Zappacosta}}},
  \bibinfo{author}{\bibfnamefont{J.~S.} \bibnamefont{{Bullock}}},
  \bibinfo{author}{\bibfnamefont{F.}~\bibnamefont{{Brighenti}}},
  \bibnamefont{and} \bibinfo{author}{\bibfnamefont{W.~G.}
  \bibnamefont{{Mathews}}}, \bibinfo{journal}{(astro-ph/0610135)}
  (\bibinfo{year}{2006}), \eprint{astro-ph/0610135}.

\bibitem[{\citenamefont{{Bolton} et~al.}(2006)\citenamefont{{Bolton}, {Burles},
  {Koopmans}, {Treu}, and {Moustakas}}}]{bolton_etal06}
\bibinfo{author}{\bibfnamefont{A.~S.} \bibnamefont{{Bolton}}},
  \bibinfo{author}{\bibfnamefont{S.}~\bibnamefont{{Burles}}},
  \bibinfo{author}{\bibfnamefont{L.~V.~E.} \bibnamefont{{Koopmans}}},
  \bibinfo{author}{\bibfnamefont{T.}~\bibnamefont{{Treu}}}, \bibnamefont{and}
  \bibinfo{author}{\bibfnamefont{L.~A.} \bibnamefont{{Moustakas}}},
  \bibinfo{journal}{\apj} \textbf{\bibinfo{volume}{638}}, \bibinfo{pages}{703}
  (\bibinfo{year}{2006}), \eprint{arXiv:astro-ph/0511453}.

\bibitem[{\citenamefont{{Johnston} et~al.}(2007)\citenamefont{{Johnston},
  {Sheldon}, {Tasitsiomi}, {Frieman}, {Wechsler}, and
  {McKay}}}]{johnston_etal07}
\bibinfo{author}{\bibfnamefont{D.~E.} \bibnamefont{{Johnston}}},
  \bibinfo{author}{\bibfnamefont{E.~S.} \bibnamefont{{Sheldon}}},
  \bibinfo{author}{\bibfnamefont{A.}~\bibnamefont{{Tasitsiomi}}},
  \bibinfo{author}{\bibfnamefont{J.~A.} \bibnamefont{{Frieman}}},
  \bibinfo{author}{\bibfnamefont{R.~H.} \bibnamefont{{Wechsler}}},
  \bibnamefont{and} \bibinfo{author}{\bibfnamefont{T.~A.}
  \bibnamefont{{McKay}}}, \bibinfo{journal}{\apj}
  \textbf{\bibinfo{volume}{656}}, \bibinfo{pages}{27} (\bibinfo{year}{2007}),
  \eprint{arXiv:astro-ph/0507467}.

\bibitem[{\citenamefont{{Gavazzi} et~al.}(2007)\citenamefont{{Gavazzi}, {Treu},
  {Rhodes}, {Koopmans}, {Bolton}, {Burles}, {Massey}, and
  {Moustakas}}}]{gavazzi_etal07}
\bibinfo{author}{\bibfnamefont{R.}~\bibnamefont{{Gavazzi}}},
  \bibinfo{author}{\bibfnamefont{T.}~\bibnamefont{{Treu}}},
  \bibinfo{author}{\bibfnamefont{J.~D.} \bibnamefont{{Rhodes}}},
  \bibinfo{author}{\bibfnamefont{L.~V.~E.} \bibnamefont{{Koopmans}}},
  \bibinfo{author}{\bibfnamefont{A.~S.} \bibnamefont{{Bolton}}},
  \bibinfo{author}{\bibfnamefont{S.}~\bibnamefont{{Burles}}},
  \bibinfo{author}{\bibfnamefont{R.}~\bibnamefont{{Massey}}}, \bibnamefont{and}
  \bibinfo{author}{\bibfnamefont{L.~A.} \bibnamefont{{Moustakas}}},
  \bibinfo{journal}{(astro-ph/0701589)}  (\bibinfo{year}{2007}),
  \eprint{astro-ph/0701589}.

\bibitem[{\citenamefont{{Comerford} and
  {Natarajan}}(2007)}]{comerford_natarajan07}
\bibinfo{author}{\bibfnamefont{J.~M.} \bibnamefont{{Comerford}}}
  \bibnamefont{and}
  \bibinfo{author}{\bibfnamefont{P.}~\bibnamefont{{Natarajan}}},
  \bibinfo{journal}{(astro-ph/0703126)}  (\bibinfo{year}{2007}),
  \eprint{astro-ph/0703126}.

\bibitem[{\citenamefont{{Hirata} and {Seljak}}(2004)}]{hirata_seljak04}
\bibinfo{author}{\bibfnamefont{C.~M.} \bibnamefont{{Hirata}}} \bibnamefont{and}
  \bibinfo{author}{\bibfnamefont{U.}~\bibnamefont{{Seljak}}},
  \bibinfo{journal}{\prd} \textbf{\bibinfo{volume}{70}},
  \bibinfo{pages}{063526} (\bibinfo{year}{2004}),
  \eprint{arXiv:astro-ph/0406275}.

\bibitem[{\citenamefont{{Heymans} et~al.}(2004)\citenamefont{{Heymans},
  {Brown}, {Heavens}, {Meisenheimer}, {Taylor}, and {Wolf}}}]{heymans_etal04}
\bibinfo{author}{\bibfnamefont{C.}~\bibnamefont{{Heymans}}},
  \bibinfo{author}{\bibfnamefont{M.}~\bibnamefont{{Brown}}},
  \bibinfo{author}{\bibfnamefont{A.}~\bibnamefont{{Heavens}}},
  \bibinfo{author}{\bibfnamefont{K.}~\bibnamefont{{Meisenheimer}}},
  \bibinfo{author}{\bibfnamefont{A.}~\bibnamefont{{Taylor}}}, \bibnamefont{and}
  \bibinfo{author}{\bibfnamefont{C.}~\bibnamefont{{Wolf}}},
  \bibinfo{journal}{\mnras} \textbf{\bibinfo{volume}{347}},
  \bibinfo{pages}{895} (\bibinfo{year}{2004}), \eprint{arXiv:astro-ph/0310174}.

\bibitem[{\citenamefont{{Mandelbaum}
  et~al.}(2006{\natexlab{b}})\citenamefont{{Mandelbaum}, {Hirata}, {Ishak},
  {Seljak}, and {Brinkmann}}}]{mandelbaum_etal06b}
\bibinfo{author}{\bibfnamefont{R.}~\bibnamefont{{Mandelbaum}}},
  \bibinfo{author}{\bibfnamefont{C.~M.} \bibnamefont{{Hirata}}},
  \bibinfo{author}{\bibfnamefont{M.}~\bibnamefont{{Ishak}}},
  \bibinfo{author}{\bibfnamefont{U.}~\bibnamefont{{Seljak}}}, \bibnamefont{and}
  \bibinfo{author}{\bibfnamefont{J.}~\bibnamefont{{Brinkmann}}},
  \bibinfo{journal}{\mnras} \textbf{\bibinfo{volume}{367}},
  \bibinfo{pages}{611} (\bibinfo{year}{2006}{\natexlab{b}}),
  \eprint{arXiv:astro-ph/0509026}.

\bibitem[{\citenamefont{{Bridle} and {King}}(2007)}]{bridle_king07}
\bibinfo{author}{\bibfnamefont{S.}~\bibnamefont{{Bridle}}} \bibnamefont{and}
  \bibinfo{author}{\bibfnamefont{L.}~\bibnamefont{{King}}},
  \bibinfo{journal}{ArXiv e-prints} \textbf{\bibinfo{volume}{705}}
  (\bibinfo{year}{2007}), \eprint{0705.0166}.

\bibitem[{\citenamefont{{Albrecht}
  et~al.}(2006{\natexlab{b}})\citenamefont{{Albrecht}, {Bernstein}, {Cahn},
  {Freedman}, {Hewitt}, {Hu}, {Huth}, {Kamionkowski}, {Kolb}, {Knox}
  et~al.}}]{detf}
\bibinfo{author}{\bibfnamefont{A.}~\bibnamefont{{Albrecht}}},
  \bibinfo{author}{\bibfnamefont{G.}~\bibnamefont{{Bernstein}}},
  \bibinfo{author}{\bibfnamefont{R.}~\bibnamefont{{Cahn}}},
  \bibinfo{author}{\bibfnamefont{W.~L.} \bibnamefont{{Freedman}}},
  \bibinfo{author}{\bibfnamefont{J.}~\bibnamefont{{Hewitt}}},
  \bibinfo{author}{\bibfnamefont{W.}~\bibnamefont{{Hu}}},
  \bibinfo{author}{\bibfnamefont{J.}~\bibnamefont{{Huth}}},
  \bibinfo{author}{\bibfnamefont{M.}~\bibnamefont{{Kamionkowski}}},
  \bibinfo{author}{\bibfnamefont{E.~W.} \bibnamefont{{Kolb}}},
  \bibinfo{author}{\bibfnamefont{L.}~\bibnamefont{{Knox}}},
  \bibnamefont{et~al.}, \bibinfo{journal}{ArXiv Astrophysics e-prints}
  (\bibinfo{year}{2006}{\natexlab{b}}), \eprint{astro-ph/0609591}.

\bibitem[{\citenamefont{{Ma} et~al.}(2006)\citenamefont{{Ma}, {Hu}, and
  {Huterer}}}]{ma_etal06}
\bibinfo{author}{\bibfnamefont{Z.}~\bibnamefont{{Ma}}},
  \bibinfo{author}{\bibfnamefont{W.}~\bibnamefont{{Hu}}}, \bibnamefont{and}
  \bibinfo{author}{\bibfnamefont{D.}~\bibnamefont{{Huterer}}},
  \bibinfo{journal}{\apj} \textbf{\bibinfo{volume}{636}}, \bibinfo{pages}{21}
  (\bibinfo{year}{2006}), \eprint{astro-ph/0506614}.

\bibitem[{\citenamefont{{Peacock} and {Dodds}}(1996)}]{peacock_dodds96}
\bibinfo{author}{\bibfnamefont{J.~A.} \bibnamefont{{Peacock}}}
  \bibnamefont{and} \bibinfo{author}{\bibfnamefont{S.~J.}
  \bibnamefont{{Dodds}}}, \bibinfo{journal}{\mnras}
  \textbf{\bibinfo{volume}{280}}, \bibinfo{pages}{L19} (\bibinfo{year}{1996}),
  \eprint{arXiv:astro-ph/9603031}.

\bibitem[{\citenamefont{{Smith} et~al.}(2003)\citenamefont{{Smith}, {Peacock},
  {Jenkins}, {White}, {Frenk}, {Pearce}, {Thomas}, {Efstathiou}, and
  {Couchman}}}]{smith_etal03}
\bibinfo{author}{\bibfnamefont{R.~E.} \bibnamefont{{Smith}}},
  \bibinfo{author}{\bibfnamefont{J.~A.} \bibnamefont{{Peacock}}},
  \bibinfo{author}{\bibfnamefont{A.}~\bibnamefont{{Jenkins}}},
  \bibinfo{author}{\bibfnamefont{S.~D.~M.} \bibnamefont{{White}}},
  \bibinfo{author}{\bibfnamefont{C.~S.} \bibnamefont{{Frenk}}},
  \bibinfo{author}{\bibfnamefont{F.~R.} \bibnamefont{{Pearce}}},
  \bibinfo{author}{\bibfnamefont{P.~A.} \bibnamefont{{Thomas}}},
  \bibinfo{author}{\bibfnamefont{G.}~\bibnamefont{{Efstathiou}}},
  \bibnamefont{and} \bibinfo{author}{\bibfnamefont{H.~M.~P.}
  \bibnamefont{{Couchman}}}, \bibinfo{journal}{\mnras}
  \textbf{\bibinfo{volume}{341}}, \bibinfo{pages}{1311} (\bibinfo{year}{2003}),
  \eprint{astro-ph/0207664}.

\bibitem[{\citenamefont{{Scherrer} and
  {Bertschinger}}(1991)}]{scherrer_bertschinger91}
\bibinfo{author}{\bibfnamefont{R.~J.} \bibnamefont{{Scherrer}}}
  \bibnamefont{and}
  \bibinfo{author}{\bibfnamefont{E.}~\bibnamefont{{Bertschinger}}},
  \bibinfo{journal}{\apj} \textbf{\bibinfo{volume}{381}}, \bibinfo{pages}{349}
  (\bibinfo{year}{1991}).

\bibitem[{\citenamefont{{Ma} and {Fry}}(2000)}]{ma_fry00}
\bibinfo{author}{\bibfnamefont{C.-P.} \bibnamefont{{Ma}}} \bibnamefont{and}
  \bibinfo{author}{\bibfnamefont{J.~N.} \bibnamefont{{Fry}}},
  \bibinfo{journal}{\apj} \textbf{\bibinfo{volume}{543}}, \bibinfo{pages}{503}
  (\bibinfo{year}{2000}), \eprint{astro-ph/0003343}.

\bibitem[{\citenamefont{{Seljak}}(2000)}]{seljak00}
\bibinfo{author}{\bibfnamefont{U.}~\bibnamefont{{Seljak}}},
  \bibinfo{journal}{\mnras} \textbf{\bibinfo{volume}{318}},
  \bibinfo{pages}{203} (\bibinfo{year}{2000}).

\bibitem[{\citenamefont{{Scoccimarro} et~al.}(2001)\citenamefont{{Scoccimarro},
  {Sheth}, {Hui}, and {Jain}}}]{scoccimarro_etal01}
\bibinfo{author}{\bibfnamefont{R.}~\bibnamefont{{Scoccimarro}}},
  \bibinfo{author}{\bibfnamefont{R.~K.} \bibnamefont{{Sheth}}},
  \bibinfo{author}{\bibfnamefont{L.}~\bibnamefont{{Hui}}}, \bibnamefont{and}
  \bibinfo{author}{\bibfnamefont{B.}~\bibnamefont{{Jain}}},
  \bibinfo{journal}{\apj} \textbf{\bibinfo{volume}{546}}, \bibinfo{pages}{20}
  (\bibinfo{year}{2001}).

\bibitem[{\citenamefont{{Sheth} et~al.}(2001)\citenamefont{{Sheth}, {Mo}, and
  {Tormen}}}]{sheth_etal01}
\bibinfo{author}{\bibfnamefont{R.~K.} \bibnamefont{{Sheth}}},
  \bibinfo{author}{\bibfnamefont{H.~J.} \bibnamefont{{Mo}}}, \bibnamefont{and}
  \bibinfo{author}{\bibfnamefont{G.}~\bibnamefont{{Tormen}}},
  \bibinfo{journal}{\mnras} \textbf{\bibinfo{volume}{323}}, \bibinfo{pages}{1}
  (\bibinfo{year}{2001}), \eprint{astro-ph/9907024}.

\bibitem[{\citenamefont{{Cooray} and {Sheth}}(2002)}]{cooray_sheth02}
\bibinfo{author}{\bibfnamefont{A.}~\bibnamefont{{Cooray}}} \bibnamefont{and}
  \bibinfo{author}{\bibfnamefont{R.}~\bibnamefont{{Sheth}}},
  \bibinfo{journal}{\physrep} \textbf{\bibinfo{volume}{372}},
  \bibinfo{pages}{1} (\bibinfo{year}{2002}).

\bibitem[{\citenamefont{{Eisenstein} and {Hu}}(1999)}]{eisenstein_hu99}
\bibinfo{author}{\bibfnamefont{D.~J.} \bibnamefont{{Eisenstein}}}
  \bibnamefont{and} \bibinfo{author}{\bibfnamefont{W.}~\bibnamefont{{Hu}}},
  \bibinfo{journal}{\apj} \textbf{\bibinfo{volume}{511}}, \bibinfo{pages}{5}
  (\bibinfo{year}{1999}), \eprint{astro-ph/9710252}.

\bibitem[{\citenamefont{{Hu}}(2002)}]{hu02}
\bibinfo{author}{\bibfnamefont{W.}~\bibnamefont{{Hu}}}, \bibinfo{journal}{\prd}
  \textbf{\bibinfo{volume}{66}}, \bibinfo{pages}{083515}
  (\bibinfo{year}{2002}), \eprint{arXiv:astro-ph/0208093}.

\bibitem[{\citenamefont{{Tinker} et~al.}(2006)\citenamefont{{Tinker},
  {Weinberg}, and {Warren}}}]{tinker_etal06}
\bibinfo{author}{\bibfnamefont{J.~L.} \bibnamefont{{Tinker}}},
  \bibinfo{author}{\bibfnamefont{D.~H.} \bibnamefont{{Weinberg}}},
  \bibnamefont{and} \bibinfo{author}{\bibfnamefont{M.~S.}
  \bibnamefont{{Warren}}}, \bibinfo{journal}{\apj}
  \textbf{\bibinfo{volume}{647}}, \bibinfo{pages}{737} (\bibinfo{year}{2006}),
  \eprint{astro-ph/0603146}.

\bibitem[{\citenamefont{{Navarro} et~al.}(1997)\citenamefont{{Navarro},
  {Frenk}, and {White}}}]{navarro_etal97}
\bibinfo{author}{\bibfnamefont{J.~F.} \bibnamefont{{Navarro}}},
  \bibinfo{author}{\bibfnamefont{C.~S.} \bibnamefont{{Frenk}}},
  \bibnamefont{and} \bibinfo{author}{\bibfnamefont{S.~D.~M.}
  \bibnamefont{{White}}}, \bibinfo{journal}{\apj}
  \textbf{\bibinfo{volume}{490}}, \bibinfo{pages}{493} (\bibinfo{year}{1997}),
  \eprint{astro-ph/9611107}.

\bibitem[{\citenamefont{{Bullock} et~al.}(2001)\citenamefont{{Bullock},
  {Kolatt}, {Sigad}, {Somerville}, {Kravtsov}, {Klypin}, {Primack}, and
  {Dekel}}}]{bullock_etal01}
\bibinfo{author}{\bibfnamefont{J.~S.} \bibnamefont{{Bullock}}},
  \bibinfo{author}{\bibfnamefont{T.~S.} \bibnamefont{{Kolatt}}},
  \bibinfo{author}{\bibfnamefont{Y.}~\bibnamefont{{Sigad}}},
  \bibinfo{author}{\bibfnamefont{R.~S.} \bibnamefont{{Somerville}}},
  \bibinfo{author}{\bibfnamefont{A.~V.} \bibnamefont{{Kravtsov}}},
  \bibinfo{author}{\bibfnamefont{A.~A.} \bibnamefont{{Klypin}}},
  \bibinfo{author}{\bibfnamefont{J.~R.} \bibnamefont{{Primack}}},
  \bibnamefont{and} \bibinfo{author}{\bibfnamefont{A.}~\bibnamefont{{Dekel}}},
  \bibinfo{journal}{\mnras} \textbf{\bibinfo{volume}{321}},
  \bibinfo{pages}{559} (\bibinfo{year}{2001}), \eprint{astro-ph/9908159}.

\bibitem[{\citenamefont{{Dolag} et~al.}(2004)\citenamefont{{Dolag},
  {Bartelmann}, {Perrotta}, {Baccigalupi}, {Moscardini}, {Meneghetti}, and
  {Tormen}}}]{dolag_etal04}
\bibinfo{author}{\bibfnamefont{K.}~\bibnamefont{{Dolag}}},
  \bibinfo{author}{\bibfnamefont{M.}~\bibnamefont{{Bartelmann}}},
  \bibinfo{author}{\bibfnamefont{F.}~\bibnamefont{{Perrotta}}},
  \bibinfo{author}{\bibfnamefont{C.}~\bibnamefont{{Baccigalupi}}},
  \bibinfo{author}{\bibfnamefont{L.}~\bibnamefont{{Moscardini}}},
  \bibinfo{author}{\bibfnamefont{M.}~\bibnamefont{{Meneghetti}}},
  \bibnamefont{and} \bibinfo{author}{\bibfnamefont{G.}~\bibnamefont{{Tormen}}},
  \bibinfo{journal}{\aap} \textbf{\bibinfo{volume}{416}}, \bibinfo{pages}{853}
  (\bibinfo{year}{2004}).

\bibitem[{\citenamefont{{Macci{\`o}} et~al.}(2007)\citenamefont{{Macci{\`o}},
  {Dutton}, {van den Bosch}, {Moore}, {Potter}, and {Stadel}}}]{maccio_etal07}
\bibinfo{author}{\bibfnamefont{A.~V.} \bibnamefont{{Macci{\`o}}}},
  \bibinfo{author}{\bibfnamefont{A.~A.} \bibnamefont{{Dutton}}},
  \bibinfo{author}{\bibfnamefont{F.~C.} \bibnamefont{{van den Bosch}}},
  \bibinfo{author}{\bibfnamefont{B.}~\bibnamefont{{Moore}}},
  \bibinfo{author}{\bibfnamefont{D.}~\bibnamefont{{Potter}}}, \bibnamefont{and}
  \bibinfo{author}{\bibfnamefont{J.}~\bibnamefont{{Stadel}}},
  \bibinfo{journal}{\mnras} \textbf{\bibinfo{volume}{378}}, \bibinfo{pages}{55}
  (\bibinfo{year}{2007}), \eprint{arXiv:astro-ph/0608157}.

\bibitem[{\citenamefont{{Wechsler} et~al.}(2006)\citenamefont{{Wechsler},
  {Zentner}, {Bullock}, {Kravtsov}, and {Allgood}}}]{wechsler_etal06}
\bibinfo{author}{\bibfnamefont{R.~H.} \bibnamefont{{Wechsler}}},
  \bibinfo{author}{\bibfnamefont{A.~R.} \bibnamefont{{Zentner}}},
  \bibinfo{author}{\bibfnamefont{J.~S.} \bibnamefont{{Bullock}}},
  \bibinfo{author}{\bibfnamefont{A.~V.} \bibnamefont{{Kravtsov}}},
  \bibnamefont{and}
  \bibinfo{author}{\bibfnamefont{B.}~\bibnamefont{{Allgood}}},
  \bibinfo{journal}{\apj} \textbf{\bibinfo{volume}{652}}, \bibinfo{pages}{71}
  (\bibinfo{year}{2006}), \eprint{astro-ph/0512416}.

\bibitem[{\citenamefont{{Neto} et~al.}(2007)\citenamefont{{Neto}, {Gao},
  {Bett}, {Cole}, {Navarro}, {Frenk}, {White}, {Springel}, and
  {Jenkins}}}]{neto_etal07}
\bibinfo{author}{\bibfnamefont{A.~F.} \bibnamefont{{Neto}}},
  \bibinfo{author}{\bibfnamefont{L.}~\bibnamefont{{Gao}}},
  \bibinfo{author}{\bibfnamefont{P.}~\bibnamefont{{Bett}}},
  \bibinfo{author}{\bibfnamefont{S.}~\bibnamefont{{Cole}}},
  \bibinfo{author}{\bibfnamefont{J.~F.} \bibnamefont{{Navarro}}},
  \bibinfo{author}{\bibfnamefont{C.~S.} \bibnamefont{{Frenk}}},
  \bibinfo{author}{\bibfnamefont{S.~D.~M.} \bibnamefont{{White}}},
  \bibinfo{author}{\bibfnamefont{V.}~\bibnamefont{{Springel}}},
  \bibnamefont{and}
  \bibinfo{author}{\bibfnamefont{A.}~\bibnamefont{{Jenkins}}},
  \bibinfo{journal}{ArXiv e-prints} \textbf{\bibinfo{volume}{706}}
  (\bibinfo{year}{2007}), \eprint{0706.2919}.

\bibitem[{\citenamefont{{Hu} and {Kravtsov}}(2003)}]{hu_kravtsov03}
\bibinfo{author}{\bibfnamefont{W.}~\bibnamefont{{Hu}}} \bibnamefont{and}
  \bibinfo{author}{\bibfnamefont{A.~V.} \bibnamefont{{Kravtsov}}},
  \bibinfo{journal}{\apj} \textbf{\bibinfo{volume}{584}}, \bibinfo{pages}{702}
  (\bibinfo{year}{2003}), \eprint{arXiv:astro-ph/0203169}.

\bibitem[{\citenamefont{{Cooray} and {Hu}}(2001)}]{cooray_hu01}
\bibinfo{author}{\bibfnamefont{A.}~\bibnamefont{{Cooray}}} \bibnamefont{and}
  \bibinfo{author}{\bibfnamefont{W.}~\bibnamefont{{Hu}}},
  \bibinfo{journal}{\apj} \textbf{\bibinfo{volume}{554}}, \bibinfo{pages}{56}
  (\bibinfo{year}{2001}), \eprint{astro-ph/0012087}.

\bibitem[{\citenamefont{{Dolney} et~al.}(2004)\citenamefont{{Dolney}, {Jain},
  and {Takada}}}]{dolney_etal04}
\bibinfo{author}{\bibfnamefont{D.}~\bibnamefont{{Dolney}}},
  \bibinfo{author}{\bibfnamefont{B.}~\bibnamefont{{Jain}}}, \bibnamefont{and}
  \bibinfo{author}{\bibfnamefont{M.}~\bibnamefont{{Takada}}},
  \bibinfo{journal}{\mnras} \textbf{\bibinfo{volume}{352}},
  \bibinfo{pages}{1019} (\bibinfo{year}{2004}),
  \eprint{arXiv:astro-ph/0401089}.

\bibitem[{\citenamefont{{Sheth} and {Tormen}}(1999)}]{sheth_tormen99}
\bibinfo{author}{\bibfnamefont{R.~K.} \bibnamefont{{Sheth}}} \bibnamefont{and}
  \bibinfo{author}{\bibfnamefont{G.}~\bibnamefont{{Tormen}}},
  \bibinfo{journal}{\mnras} \textbf{\bibinfo{volume}{308}},
  \bibinfo{pages}{119} (\bibinfo{year}{1999}), \eprint{astro-ph/9901122}.

\bibitem[{\citenamefont{{White} and {Hu}}(2000)}]{white_hu00}
\bibinfo{author}{\bibfnamefont{M.}~\bibnamefont{{White}}} \bibnamefont{and}
  \bibinfo{author}{\bibfnamefont{W.}~\bibnamefont{{Hu}}},
  \bibinfo{journal}{\apj} \textbf{\bibinfo{volume}{537}}, \bibinfo{pages}{1}
  (\bibinfo{year}{2000}).

\bibitem[{\citenamefont{{Vale} and {White}}(2003)}]{vale_white03}
\bibinfo{author}{\bibfnamefont{C.}~\bibnamefont{{Vale}}} \bibnamefont{and}
  \bibinfo{author}{\bibfnamefont{M.}~\bibnamefont{{White}}},
  \bibinfo{journal}{Apj} \textbf{\bibinfo{volume}{592}}, \bibinfo{pages}{699}
  (\bibinfo{year}{2003}).

\bibitem[{\citenamefont{{Dodelson} et~al.}(2006)\citenamefont{{Dodelson},
  {Shapiro}, and {White}}}]{dodelson_etal06}
\bibinfo{author}{\bibfnamefont{S.}~\bibnamefont{{Dodelson}}},
  \bibinfo{author}{\bibfnamefont{C.}~\bibnamefont{{Shapiro}}},
  \bibnamefont{and} \bibinfo{author}{\bibfnamefont{M.}~\bibnamefont{{White}}},
  \bibinfo{journal}{\prd} \textbf{\bibinfo{volume}{73}},
  \bibinfo{pages}{023009} (\bibinfo{year}{2006}),
  \eprint{arXiv:astro-ph/0508296}.

\bibitem[{\citenamefont{{Semboloni} et~al.}(2007)\citenamefont{{Semboloni},
  {van Waerbeke}, {Heymans}, {Hamana}, {Colombi}, {White}, and
  {Mellier}}}]{semboloni_etal06}
\bibinfo{author}{\bibfnamefont{E.}~\bibnamefont{{Semboloni}}},
  \bibinfo{author}{\bibfnamefont{L.}~\bibnamefont{{van Waerbeke}}},
  \bibinfo{author}{\bibfnamefont{C.}~\bibnamefont{{Heymans}}},
  \bibinfo{author}{\bibfnamefont{T.}~\bibnamefont{{Hamana}}},
  \bibinfo{author}{\bibfnamefont{S.}~\bibnamefont{{Colombi}}},
  \bibinfo{author}{\bibfnamefont{M.}~\bibnamefont{{White}}}, \bibnamefont{and}
  \bibinfo{author}{\bibfnamefont{Y.}~\bibnamefont{{Mellier}}},
  \bibinfo{journal}{\mnras} \textbf{\bibinfo{volume}{375}}, \bibinfo{pages}{L6}
  (\bibinfo{year}{2007}), \eprint{arXiv:astro-ph/0606648}.

\bibitem[{\citenamefont{{Huterer} and {Turner}}(2001)}]{huterer_turner01}
\bibinfo{author}{\bibfnamefont{D.}~\bibnamefont{{Huterer}}} \bibnamefont{and}
  \bibinfo{author}{\bibfnamefont{M.~S.} \bibnamefont{{Turner}}},
  \bibinfo{journal}{\prd} \textbf{\bibinfo{volume}{64}},
  \bibinfo{pages}{123527} (\bibinfo{year}{2001}),
  \eprint{arXiv:astro-ph/0012510}.

\bibitem[{\citenamefont{{Hu} and {Jain}}(2004)}]{hu_jain04}
\bibinfo{author}{\bibfnamefont{W.}~\bibnamefont{{Hu}}} \bibnamefont{and}
  \bibinfo{author}{\bibfnamefont{B.}~\bibnamefont{{Jain}}},
  \bibinfo{journal}{\prd} \textbf{\bibinfo{volume}{70}},
  \bibinfo{pages}{043009} (\bibinfo{year}{2004}),
  \eprint{arXiv:astro-ph/0312395}.

\bibitem[{\citenamefont{{Katz} and {White}}(1993)}]{katz_white93}
\bibinfo{author}{\bibfnamefont{N.}~\bibnamefont{{Katz}}} \bibnamefont{and}
  \bibinfo{author}{\bibfnamefont{S.~D.~M.} \bibnamefont{{White}}},
  \bibinfo{journal}{\apj} \textbf{\bibinfo{volume}{412}}, \bibinfo{pages}{455}
  (\bibinfo{year}{1993}).

\bibitem[{\citenamefont{{Lewis} et~al.}(2000)\citenamefont{{Lewis}, {Babul},
  {Katz}, {Quinn}, {Hernquist}, and {Weinberg}}}]{lewis_etal00}
\bibinfo{author}{\bibfnamefont{G.~F.} \bibnamefont{{Lewis}}},
  \bibinfo{author}{\bibfnamefont{A.}~\bibnamefont{{Babul}}},
  \bibinfo{author}{\bibfnamefont{N.}~\bibnamefont{{Katz}}},
  \bibinfo{author}{\bibfnamefont{T.}~\bibnamefont{{Quinn}}},
  \bibinfo{author}{\bibfnamefont{L.}~\bibnamefont{{Hernquist}}},
  \bibnamefont{and} \bibinfo{author}{\bibfnamefont{D.~H.}
  \bibnamefont{{Weinberg}}}, \bibinfo{journal}{\apj}
  \textbf{\bibinfo{volume}{536}}, \bibinfo{pages}{623} (\bibinfo{year}{2000}),
  \eprint{astro-ph/9907097}.

\bibitem[{\citenamefont{{Pearce} et~al.}(2000)\citenamefont{{Pearce}, {Thomas},
  {Couchman}, and {Edge}}}]{pearce_etal00}
\bibinfo{author}{\bibfnamefont{F.~R.} \bibnamefont{{Pearce}}},
  \bibinfo{author}{\bibfnamefont{P.~A.} \bibnamefont{{Thomas}}},
  \bibinfo{author}{\bibfnamefont{H.~M.~P.} \bibnamefont{{Couchman}}},
  \bibnamefont{and} \bibinfo{author}{\bibfnamefont{A.~C.}
  \bibnamefont{{Edge}}}, \bibinfo{journal}{\mnras}
  \textbf{\bibinfo{volume}{317}}, \bibinfo{pages}{1029} (\bibinfo{year}{2000}),
  \eprint{astro-ph/9908062}.

\bibitem[{\citenamefont{{Dav{\'e}} et~al.}(2001)\citenamefont{{Dav{\'e}},
  {Cen}, {Ostriker}, {Bryan}, {Hernquist}, {Katz}, {Weinberg}, {Norman}, and
  {O'Shea}}}]{dave_etal01}
\bibinfo{author}{\bibfnamefont{R.}~\bibnamefont{{Dav{\'e}}}},
  \bibinfo{author}{\bibfnamefont{R.}~\bibnamefont{{Cen}}},
  \bibinfo{author}{\bibfnamefont{J.~P.} \bibnamefont{{Ostriker}}},
  \bibinfo{author}{\bibfnamefont{G.~L.} \bibnamefont{{Bryan}}},
  \bibinfo{author}{\bibfnamefont{L.}~\bibnamefont{{Hernquist}}},
  \bibinfo{author}{\bibfnamefont{N.}~\bibnamefont{{Katz}}},
  \bibinfo{author}{\bibfnamefont{D.~H.} \bibnamefont{{Weinberg}}},
  \bibinfo{author}{\bibfnamefont{M.~L.} \bibnamefont{{Norman}}},
  \bibnamefont{and} \bibinfo{author}{\bibfnamefont{B.}~\bibnamefont{{O'Shea}}},
  \bibinfo{journal}{\apj} \textbf{\bibinfo{volume}{552}}, \bibinfo{pages}{473}
  (\bibinfo{year}{2001}), \eprint{astro-ph/0007217}.

\bibitem[{\citenamefont{{Balogh} et~al.}(2001)\citenamefont{{Balogh}, {Pearce},
  {Bower}, and {Kay}}}]{balogh_etal01}
\bibinfo{author}{\bibfnamefont{M.~L.} \bibnamefont{{Balogh}}},
  \bibinfo{author}{\bibfnamefont{F.~R.} \bibnamefont{{Pearce}}},
  \bibinfo{author}{\bibfnamefont{R.~G.} \bibnamefont{{Bower}}},
  \bibnamefont{and} \bibinfo{author}{\bibfnamefont{S.~T.} \bibnamefont{{Kay}}},
  \bibinfo{journal}{\mnras} \textbf{\bibinfo{volume}{326}},
  \bibinfo{pages}{1228} (\bibinfo{year}{2001}), \eprint{astro-ph/0104041}.

\bibitem[{\citenamefont{{Borgani} et~al.}(2002)\citenamefont{{Borgani},
  {Governato}, {Wadsley}, {Menci}, {Tozzi}, {Quinn}, {Stadel}, and
  {Lake}}}]{borgani_etal02}
\bibinfo{author}{\bibfnamefont{S.}~\bibnamefont{{Borgani}}},
  \bibinfo{author}{\bibfnamefont{F.}~\bibnamefont{{Governato}}},
  \bibinfo{author}{\bibfnamefont{J.}~\bibnamefont{{Wadsley}}},
  \bibinfo{author}{\bibfnamefont{N.}~\bibnamefont{{Menci}}},
  \bibinfo{author}{\bibfnamefont{P.}~\bibnamefont{{Tozzi}}},
  \bibinfo{author}{\bibfnamefont{T.}~\bibnamefont{{Quinn}}},
  \bibinfo{author}{\bibfnamefont{J.}~\bibnamefont{{Stadel}}}, \bibnamefont{and}
  \bibinfo{author}{\bibfnamefont{G.}~\bibnamefont{{Lake}}},
  \bibinfo{journal}{\mnras} \textbf{\bibinfo{volume}{336}},
  \bibinfo{pages}{409} (\bibinfo{year}{2002}), \eprint{astro-ph/0205471}.

\bibitem[{\citenamefont{{Suginohara} and
  {Ostriker}}(1998)}]{suginohara_ostriker98}
\bibinfo{author}{\bibfnamefont{T.}~\bibnamefont{{Suginohara}}}
  \bibnamefont{and} \bibinfo{author}{\bibfnamefont{J.~P.}
  \bibnamefont{{Ostriker}}}, \bibinfo{journal}{\apj}
  \textbf{\bibinfo{volume}{507}}, \bibinfo{pages}{16} (\bibinfo{year}{1998}),
  \eprint{astro-ph/9803318}.

\bibitem[{\citenamefont{Cooray et~al.}(2000)\citenamefont{Cooray, Hu, and
  Miralda-Escude}}]{CooHuMir00}
\bibinfo{author}{\bibfnamefont{A.}~\bibnamefont{Cooray}},
  \bibinfo{author}{\bibfnamefont{W.}~\bibnamefont{Hu}}, \bibnamefont{and}
  \bibinfo{author}{\bibfnamefont{J.}~\bibnamefont{Miralda-Escude}},
  \bibinfo{journal}{Astrophys. J.} \textbf{\bibinfo{volume}{535}},
  \bibinfo{pages}{L9} (\bibinfo{year}{2000}), \eprint{astro-ph/0003205}.

\bibitem[{\citenamefont{{Bahcall} et~al.}(2000)\citenamefont{{Bahcall}, {Cen},
  {Dav{\'e}}, {Ostriker}, and {Yu}}}]{bahcall_etal00}
\bibinfo{author}{\bibfnamefont{N.~A.} \bibnamefont{{Bahcall}}},
  \bibinfo{author}{\bibfnamefont{R.}~\bibnamefont{{Cen}}},
  \bibinfo{author}{\bibfnamefont{R.}~\bibnamefont{{Dav{\'e}}}},
  \bibinfo{author}{\bibfnamefont{J.~P.} \bibnamefont{{Ostriker}}},
  \bibnamefont{and} \bibinfo{author}{\bibfnamefont{Q.}~\bibnamefont{{Yu}}},
  \bibinfo{journal}{\apj} \textbf{\bibinfo{volume}{541}}, \bibinfo{pages}{1}
  (\bibinfo{year}{2000}), \eprint{arXiv:astro-ph/0002310}.

\bibitem[{\citenamefont{{Tinker} et~al.}(2005)\citenamefont{{Tinker},
  {Weinberg}, {Zheng}, and {Zehavi}}}]{tinker_etal05}
\bibinfo{author}{\bibfnamefont{J.~L.} \bibnamefont{{Tinker}}},
  \bibinfo{author}{\bibfnamefont{D.~H.} \bibnamefont{{Weinberg}}},
  \bibinfo{author}{\bibfnamefont{Z.}~\bibnamefont{{Zheng}}}, \bibnamefont{and}
  \bibinfo{author}{\bibfnamefont{I.}~\bibnamefont{{Zehavi}}},
  \bibinfo{journal}{\apj} \textbf{\bibinfo{volume}{631}}, \bibinfo{pages}{41}
  (\bibinfo{year}{2005}), \eprint{astro-ph/0411777}.

\bibitem[{\citenamefont{{van den Bosch} et~al.}(2007)\citenamefont{{van den
  Bosch}, {Yang}, {Mo}, {Weinmann}, {Macci{\`o}}, {More}, {Cacciato}, {Skibba},
  and {Kang}}}]{vdb_etal07}
\bibinfo{author}{\bibfnamefont{F.~C.} \bibnamefont{{van den Bosch}}},
  \bibinfo{author}{\bibfnamefont{X.}~\bibnamefont{{Yang}}},
  \bibinfo{author}{\bibfnamefont{H.~J.} \bibnamefont{{Mo}}},
  \bibinfo{author}{\bibfnamefont{S.~M.} \bibnamefont{{Weinmann}}},
  \bibinfo{author}{\bibfnamefont{A.~V.} \bibnamefont{{Macci{\`o}}}},
  \bibinfo{author}{\bibfnamefont{S.}~\bibnamefont{{More}}},
  \bibinfo{author}{\bibfnamefont{M.}~\bibnamefont{{Cacciato}}},
  \bibinfo{author}{\bibfnamefont{R.}~\bibnamefont{{Skibba}}}, \bibnamefont{and}
  \bibinfo{author}{\bibfnamefont{X.}~\bibnamefont{{Kang}}},
  \bibinfo{journal}{\mnras} \textbf{\bibinfo{volume}{376}},
  \bibinfo{pages}{841} (\bibinfo{year}{2007}), \eprint{arXiv:astro-ph/0610686}.

\bibitem[{\citenamefont{{Spergel} et~al.}(2003)\citenamefont{{Spergel},
  {Verde}, {Peiris}, {Komatsu}, {Nolta}, {Bennett}, {Halpern}, {Hinshaw},
  {Jarosik}, {Kogut} et~al.}}]{spergel_etal03}
\bibinfo{author}{\bibfnamefont{D.~N.} \bibnamefont{{Spergel}}},
  \bibinfo{author}{\bibfnamefont{L.}~\bibnamefont{{Verde}}},
  \bibinfo{author}{\bibfnamefont{H.~V.} \bibnamefont{{Peiris}}},
  \bibinfo{author}{\bibfnamefont{E.}~\bibnamefont{{Komatsu}}},
  \bibinfo{author}{\bibfnamefont{M.~R.} \bibnamefont{{Nolta}}},
  \bibinfo{author}{\bibfnamefont{C.~L.} \bibnamefont{{Bennett}}},
  \bibinfo{author}{\bibfnamefont{M.}~\bibnamefont{{Halpern}}},
  \bibinfo{author}{\bibfnamefont{G.}~\bibnamefont{{Hinshaw}}},
  \bibinfo{author}{\bibfnamefont{N.}~\bibnamefont{{Jarosik}}},
  \bibinfo{author}{\bibfnamefont{A.}~\bibnamefont{{Kogut}}},
  \bibnamefont{et~al.}, \bibinfo{journal}{\apjs}
  \textbf{\bibinfo{volume}{148}}, \bibinfo{pages}{175} (\bibinfo{year}{2003}),
  \eprint{arXiv:astro-ph/0302209}.

\bibitem[{\citenamefont{{Huterer} and {White}}(2005)}]{huterer_white05}
\bibinfo{author}{\bibfnamefont{D.}~\bibnamefont{{Huterer}}} \bibnamefont{and}
  \bibinfo{author}{\bibfnamefont{M.}~\bibnamefont{{White}}},
  \bibinfo{journal}{\prd} \textbf{\bibinfo{volume}{72}},
  \bibinfo{pages}{043002} (\bibinfo{year}{2005}),
  \eprint{arXiv:astro-ph/0501451}.

\bibitem[{\citenamefont{{Hu} et~al.}(2006)\citenamefont{{Hu}, {Huterer}, and
  {Smith}}}]{hu_etal06}
\bibinfo{author}{\bibfnamefont{W.}~\bibnamefont{{Hu}}},
  \bibinfo{author}{\bibfnamefont{D.}~\bibnamefont{{Huterer}}},
  \bibnamefont{and} \bibinfo{author}{\bibfnamefont{K.~M.}
  \bibnamefont{{Smith}}}, \bibinfo{journal}{\apjl}
  \textbf{\bibinfo{volume}{650}}, \bibinfo{pages}{L13} (\bibinfo{year}{2006}),
  \eprint{arXiv:astro-ph/0607316}.

\bibitem[{\citenamefont{Bernstein and Jain}(2004)}]{Bernstein:2003es}
\bibinfo{author}{\bibfnamefont{G.~M.} \bibnamefont{Bernstein}}
  \bibnamefont{and} \bibinfo{author}{\bibfnamefont{B.}~\bibnamefont{Jain}},
  \bibinfo{journal}{Astrophys. J.} \textbf{\bibinfo{volume}{600}},
  \bibinfo{pages}{17} (\bibinfo{year}{2004}), \eprint{astro-ph/0309332}.

\end{thebibliography}

\end{document}